\documentclass[12pt]{article}

\usepackage{amsfonts,amsmath,amssymb,bm,graphicx,hyperref,a4wide,cite,mathabx,subfigure,natbib}

\numberwithin{equation}{section}

\begin{document}

\title{\textbf{Magnetic helicity in rotating neutron stars}}

\author{Maxim Dvornikov\thanks{maxdvo@izmiran.ru} 
\\
\small{\ Pushkov Institute of Terrestrial Magnetism, Ionosphere} \\
\small{and Radiowave Propagation (IZMIRAN),} \\
\small{108840 Troitsk, Moscow, Russia}}

\date{}

\maketitle

\begin{abstract}
We study the magnetic helicity evolution in neutron stars (NSs). First, we analyze how the surface terms affect the conservation law for the sum of the chiral imbalance of the charged particle densities and the density of the magnetic helicity. Our results are applied to a system with a finite volume, which can be a magnetized NS. We show that the contribution of these surface terms can potentially lead to the reconnection of magnetic field lines followed by X-ray or gamma bursts observed in magnetars. Comparing the new quantum surface term with the classical contribution known in the standard magnetic hydrodynamics, we obtain that its contribution dominates over the classical term only for  NS with a rigid rotation. Second, we study the dynamics of chiral electrons which interact electroweakly with a background
fermions having the velocity, arbitrarily
depending on spatial coordinates, and the nonuniform density. We derive the the kinetic equations and the effective action for right and left particles. The correction to the Adler anomaly and to the anomalous electric current are obtained in the case of a rotating matter. Then, the obtained results are applied for the study of the magnetic helicity flow inside a magnetized rotating NS.
We compute the characteristic time of the helicity change and demonstrate that it coincides with the the magnetic cycle period of certain pulsars.
\end{abstract}

\section{Introduction}

Studies of various phenomena taking place in the systems of chiral fermions, apart from a solely theoretical
interest, have various applications in astroparticle physics, cosmology, solid state and
accelerator physics. By chiral media we mean the situations when either the temperature or the chemical potential being much greater than the particle mass.

One should mention, first, the chiral magnetic effect (the CME), which is the generation the anomalous electric current along an external magnetic field. Such a current can exist only in matter with a nonzero chiral imbalance, i.e. the chemical potentials of left and right fermions should be different. We also recall that an ultrarelativistic fermion has a certain projection of its spin $\mathbf{s}$ on the particle momentum $\mathbf{p}$. If $\mathbf{s} \upuparrows \mathbf{p}$, we speak about right particles. The opposite situation of $\mathbf{s} \updownarrows \mathbf{p}$ corresponds to left fermions. The existence of an anomalous current $\mathbf{j}\parallel \mathbf{B}$ in vacuum is forbidden by the parity conservation in electrodynamics. The CME was predicted first by~\citet{Vil80a} and, then, reconsidered by~\citet{Fuk08}.

The next important chiral phenomenon to mention is the chiral vortical effect (the CVE). It is the excitation of an electric current along the vorticity of matter in a rotating system of ultrarelativistic fermions. The CVE is also forbidden in vacuum. The existence of the CVE was also predicted theoretically by~\citet{Vil80b}. We also recall the chiral separation effect (the CSE) which consists in the production of an axial current along an external magnetic field. Other interesting chiral phenomena are summarized in the review by~\citet{Kha16}.

\citet{Kha16,ZhaWan19} suggested that one can explain
the asymmetry in collisions of heavy ions on the basis of the CME and the CVE.
The CME and the CSE were reported to be
observed in magnetized
atomic gases~\citep{Hua16} and in Weyl and Dirac semimetals~\citep{Arm18}.

There are numerous applications of the chiral phenomena in cosmology. For instance, \citet{BoyFroRuc12} showed that the evolution of magnetic fields in a primordial plasma is strongly affected by the quantum chiral anomaly at temperatures $T > 10\,\text{MeV}$. \citet{BoyFroRuc12} also found that the magnetic helicity in plasma of the early universe after
the electroweak phase transition becomes long scale. This process leads to the production of the lepton asymmetry in the primordial plasma cooling down to $T \sim 10\,\text{MeV}$. \citet{GioSha98} assumed that a plasma before the electroweak phase transition ($T>100\,\text{GeV}$) can be permeated by a primordial hypermagnetic field. This field can be transformed to the fermionic number owing to the anomalous coupling. The influence of primordial magnetic fields on the electroweak phase transition was analyzed. The production of the baryon asymmetry in some specific configurations of the magnetic field was studied.

There are applications of the CME and the CVE in astrophysics. They are related to the possible explanation
of the strong magnetic fields existence in some compact stars. \citet{KasBel17} showed that magnetic fields with the strength $B>10^{15}\,\text{G}$ are observed in magnetars. \citet{Yam16} accounted for the chirality of neutrinos in the conventional hydrodynamics and the kinetic theory. It was suggested that a dense gas of chiral neutrinos can be important in generation of a strong helical magnetic field in a core-collapsing supernova. \citet{Yam16} provided a mechanism for the transformation of the gravitational energy of a collapsing star to the electromagnetic energy and explained strong magnetic fields in magnetars. \citet{Mas18} used the chiral magnetohydrodynamics to describe the macroscopic evolution of relativistic charged matter with a chiral imbalance. Real-time evolutions of magnetic fields and the properties of the chiral MHD turbulence were studied numerically. The inverse cascade of the magnetic energy was observed. The application of the results for the generation of strong magnetic fields in magnetars was discussed by~\citet{Mas18}. \citet{Kam16} studied the neutrino emission along the vorticity or the magnetic field in dense matter of a neutron star (NS) taking into account for the effects of the anomalous hydrodynamics. The results were applied to account for the great linear velocities $\sim 10^3\,\text{km}\cdot\text{s}^{-1}$ observed of some pulsars. \citet{SigLei16} examined the effect of the chiral instability of the spectra of the magnetic energy density and the helicity density during the early cooling phase of a hot protoneutron star. The generation of a magnetic field with the strength close to that observed in magnetars was predicted. The density fluctuations and the turbulence spectrum of matter were found to affect the generation of the magnetic field. The scale of the magnetic field was obtained to be from the submillimeter to cm range.

\citet{DvoSem15,DvoSem15b,DvoSem15c} suggested that the instability of a magnetic field caused by the electroweak correction to the CME leads to the generation of strongs fields observed in magnetars. Analogous mechanism was used by~\citet{Dvo16a,Dvo18} in quark stars. \citet{Dvo20} accounted for the electroweak interaction between chiral electrons and background matter in a rotating NS. \citet{DvoSem18b,DvoSem19} studied the evolution of the magnetic helicity and the magnetic field in NS with inhonogeneous matter using the Anomalous MagnetoHydroDynamics (AMHD) and applied the results for various problems of magnetars. \citet{DvoSemSok20} applied the CME and the CSE in a nonuniformly heated medium of a protoneutron star to explain strong magnetic fields of magnetars.

In the
present work, we continue our previous research on the evolution of magnetic fields in compact stars.
%Now, we analyze
%how the evolution of the magnetic helicity is driven by the electroweak
%interaction of chiral fermions with dense inhomogeneous electroweak
%matter inside a compact star.
We tackle two main problems. First, we examine how AMHD is modified when we account for the presence of a nonzero electron mass in a magnetized plasma. We recall that one assumes the zero electron mass, $m=0$, to support the CME based on the anomalous current~\citep{Fuk08} (see Eqs.~(\ref{anomalous_current}) and~(\ref{CME} below, as well as works by~\citet{BoyFroRuc12,Bra17}) in the AMHD.

Note that the particles are massive, $m_e\neq 0$, in realistic cosmological and astrophysical media, e.g., in the ultrarelativistic degenerate electron gas in a supernova, where $p_{\mathrm{F}_e}\gg m_e$, or in a hot primordial plasma of the early universe with $T\gg m_e$. \citet{Dvo16b} demonstrated that it results in a very rapid washing out of a chiral imbalance, $\mu_5\to 0$, because of the spin-flip in particle collisions.

If we take into account the nonzero mass $m_e\neq 0$ for left and right electrons or positrons, the CME is no longer valid. Nevertheless, \citet{Iof06} showed that the Adler-Bell-Jackiw anomaly still takes place since it is valid for massive particles as well. Thus, the standard MHD and the AMHD are compatible. It happens under the following additional condition: a change of the magnetic helicity within a domain volume is given by the new quantum effect consisting in the nonzero flux of the averaged spin through the surface of that domain. The evolution of the magnetic helicity accounts for this new term which can be comparable with the known helicity losses at the surface within classical MHD~\citep{Pri16}. We recall that
the evolution of the magnetic helicity at a stellar surface is important for the reconnection of magnetic field lines there. It can lead to the efficient conversion of the magnetic energy into the kinetic and thermal energies of plasma causing a strong electromagnetic emission of magnetars~\citep{KasBel17}.

In the present work, we also study the behavior of ultrarelativistic electrons in plasma accounting for their electroweak interaction with nonrelativistic background matter composed of protons and neutrons. The electroweak interaction is treated within the Fermi approximation. We suppose that the characteristics of this matter, such as the macroscopic velocity and the number density can depend on spatial coordinates. We take that the electron gas is degenerate, i.e. the Fermi momentum of electrons $p_{\mathrm{F}e}$ is much greater than the mass of an electron $m_e$. The existence of such a system is possible in NS. One can apply the chiral phenomena to such fermions since these particles are ultrarelativistic: $p_{\mathrm{F}e}\gg m_e$.
The mixing between left and right fermions, mentioned above, depends on the spin-flip rate in particle collisions.
Anomalous electric currents of protons are negligible since protons are nonrelativistic in NS. Hydrodynamic currents of neutrinos, which are ultrarelativistic particles, can receive an anomalous contribution. However, the fluxes of neutrinos are small for an old NS, studied here. Thus, we neglect the neutrino contribution.

This work is organized in the following way.
%In the present paper, we summarize the results of~\citet{DvoSem18b,Dvo20}.
First, in Sec.~\ref{sec:MAGNETIZ}, following~\citet{DvoSem18b}, we remind some known results for the spin magnetization in plasma and some formulas describing the electron density in the relativistic degenerate electron gas of a magnetized NS. Then, in Sec.~\ref{sec:QUANTCORR}, we derive from the Adler-Bell-Jackiw anomaly the surface terms for the modified CME accounting for all spin terms including that originated by the pseudoscalar ($\sim \bar{\psi}\gamma_5\psi$). Here $\gamma_5 = \mathrm{i}\gamma^0\gamma^1\gamma^2\gamma^3$ and $\gamma^\mu = (\gamma^0,\bm{\gamma})$ are the Dirac matrices. In Sec.~\ref{sec:HELDISS}, we derive the new magnetic helicity evolution equation, where, neglecting chiral anomaly $n_\mathrm{R} - n_\mathrm{L}=0$ due to the spin-flip, we find the contribution of the mean spin flux through the boundary of the volume to the dynamics of the magnetic helicity known in classical MHD.
In Sec.~\ref{sec:HELEVOL}, we compare estimates of magnetic helicity losses given by the classical MHD~\citep{Pri16} and our new (quantum) contribution given by that mean spin flux. 

Then, in Sec.~\ref{sec:EVOL}, basing on the results of~\citet{Dvo20},
we formulate the dynamics of chiral fermions in dense inhomogeneous
matter, with the electroweak interaction being accounted for. The
evolution of chiral fermions was described approximately basing on
the Berry phase evolution. We find the effective actions and the kinetic
equations for chiral fermions in Sec.~\ref{sec:EVOL}. Then, in Sec.~\ref{sec:ELCURR},
we derive the corrections to the anomalous electric current and to
the Adler anomaly from the electroweak interaction with inhomogeneous
matter. Finally, we apply our results in Sec.~\ref{sec:APPL} for
the description of the magnetic helicity evolution in a rotating NS and compare our prediction for the period of the magnetic
cycle with the astronomical observations. We conclude in Sec.~\ref{sec:CONCL}.

\section{Theoretical background\label{sec:THBACKGR}}

In this section, we summarize the theoretical background necessary for the application of the chiral phenomena to the evolution of magnetic fields in NSs. First, in Sec.~\ref{sec:MAGNETIZ}, we calculate the mean spin of relativistic electron gas in an external magnetic field. Using the Berry phase approximation in Sec.~\ref{sec:EVOL}, we derive the kinetic equation for electrons in rotating matter accounting for the electroweak interaction between fermions. Basing on the results of Sec.~\ref{sec:EVOL}, we derive the correction to the electic currents and to the Alder anomaly in Sec.~\ref{sec:ELCURR}.

\subsection{Mean spin of the electron gas in an external magnetic field\label{sec:MAGNETIZ}}
%\vskip0.3cm

Here, we remind the basic properties of plasma in an external magnetic field. The motion of a relativistic charged particle in a magnetic field obeys the Dirac equation. The energy levels of a $1/2$-spin fermion with the electric charge $q$ in an external magnetic field ${\bf H}=(0,0,H)$ has the form~\citep[pp.~121--122]{BerLifPit82},
\begin{equation}\label{levels}
  \mathcal{E}(p_z, n,\lambda)  =
  \sqrt{m_e^2 +p_z^2 + |q| H (2n+ 1) - q H \lambda}.
\end{equation}
where $p_z$ is the conserved projection of the fermion momentum along the magnetic field, $n = 0,1,\dots$ is the discrete main quantum number, and $\lambda = \pm 1$ is the eigenvalue of the matrix $\Sigma_z = \gamma^5\gamma^0\gamma^3$, which appears in the squared Dirac equation.\footnote{We remind that the commutator of the matrix $\Sigma_z$ with the Hamiltonian is nonzero for a moving Dirac particle~\cite[p.~86]{BerLifPit82}. However, we shall use the quantity $\lambda$ in the energy levels in Eq.~\eqref{levels}.} Negatively charged particles (electrons) should have $q=-e$ and positively charged ones (positrons) possess $q=+e$. Here $e>0$ is the absolute value of the elementary charge.

%Let us remind some known formulas.
%The signs of dubbled spin projections on the magnetic field ${\bf H}=(0,0,H)$ entering the paramagnetic terms for electrons and positrons at the main Landau level $n=0$, are opposite, i.e.. $\lambda=\mp 1$ in the known formula for energy levels (upper sign for electrons, lower for positrons),
%\begin{equation}\label{levels}
%\varepsilon_p(p_z, n,\lambda)=\sqrt{m_e^2 +p_z^2 + eH(2n+ 1 \pm \lambda)}.
%\end{equation}

Let us consider $e^{\pm}$ plasma in the external magnetic field. \citet{SV} found that the main Landau level with $n=0$ contributes the mean spin of electrons and positrons, which has the form,
%
%Since the  main Landau level $n=0$ contributes to the mean spin only,.
%accounting for the operator reordering $\hat{d}\hat{d}^+= - \hat{d}^+\hat{d}$ for positrons, one obtains the magnetization in $e^{\pm}$ plasma \cite{SV}
%
%\begin{eqnarray}\label{total}
%&&M_j^{(e)} - M_j^{(\bar{e})}=\mu_\mathrm{B}\langle \bar{\psi}_e\gamma_j\gamma_5\psi_e\rangle_{0}=\mu_\mathrm{B}\langle \psi_e^+\Sigma_j\psi_e\rangle_{0}=\frac{2e\mu_\mathrm{B}H_j}{(2\pi)^2}\nonumber\\&&\times\int_0^{\infty}dp_z\left[\frac{1}{\exp[(\sqrt{p_z^2 + m_e^2} -\mu_e)/T] + 1} -\frac{1}{\exp[(\sqrt{p_z^2 + m_e^2} +\mu_e)/T] + 1}\right],\nonumber\\
%\end{eqnarray}
%
\begin{align}\label{total}
  \bm{\mathcal{S}} = & 
  \langle \psi_e^\dagger \bm{\Sigma} \psi_e\rangle_{0} =
  - \frac{2e\mathbf{H}}{(2\pi)^2}
  \notag
  \\
  & \times
  \int_0^{\infty}\mathrm{d}p_z
  \left\{
    \frac{1}{\exp[(\sqrt{p_z^2 + m_e^2} -\mu_e)/T] + 1}-
    \frac{1}{\exp[(\sqrt{p_z^2 + m_e^2} +\mu_e)/T] + 1}
  \right\},
\end{align}
where $\psi_e$ is the bispinor exactly accounting for the contribution of the external magnetic field, $\bm{\Sigma} = \gamma^5\gamma^0\bm{\gamma}$,
%$\mu_\mathrm{B}=e/2m_e>0$ is the Bohr magneton,
$\mu_e$ is the chemical potential of relativistic electrons, and $T$ is the temperature of plasma.
%In Eq.~\eqref{total}, we take into account the fact that the magnetic moment of a electron is negative: $-\mu_\mathrm{B}$.
 
We neglect the electron mass in a chiral plasma, $m_e\to 0$. Thus, the integral in the last line is proportional to the chemical potential $\mu_e$. This result does not dependent on the temperature. Therefore the mean spin in of relativistic electron-positron plasma 
%(normalized on $\mu_\mathrm{B}$)
takes the form,
\begin{equation}\label{chiral-limit}
  \bm{\mathcal{S}} = -
  \frac{e\mu_e{\bf H}}{2\pi^2}= -
  \left(
    \frac{\alpha_\mathrm{em}}{\pi}
  \right)
  \left(
    \frac{2\mu_e}{e}
  \right)
  {\bf H},
\end{equation}
where $\alpha_\mathrm{em}=e^2/4\pi\approx 7.3\times 10^{-3}$ is the fine structure constant.

We can also consider a non-relativistic degenerate electron gas, e.g., forming plasma in a metal. In this situation, using the chemical potential $\mu_e=m_e + p^2_{\mathrm{F}_e}/2m_e$ in Eq.~(\ref{total}), one obtains the paramagnetic magnetization term $\mathbf{M}=-\mu_\mathrm{B}\bm{\mathcal{S}}$,\footnote{Here we account for the negative magnetic moment of an electron: $-\mu_\mathrm{B}$.}
\begin{equation}\label{NRplasma}
  \frac{\mathbf{M}}{\mu_\mathrm{B}}=\frac{ep_{\mathrm{F}_e}}{2\pi^2}{\bf H},
\end{equation}
where $\mu_\mathrm{B}=e/2m_e>0$ is the Bohr magneton.
This result means that the time independent paramagnetic susceptibility is equal to the known value $\chi=\alpha_\mathrm{em}v_{\mathrm{F}_e}/\pi\ll 1$ where  $v_{\mathrm{F}_e}=p_{\mathrm{F}_e}/m_e$ is the Fermi velocity. It leads to the standard definitions $\mathbf{M}=\chi \mathbf{H}$ and $\mathbf{B}=\mu \mathbf{H}= (1 + \chi ) \mathbf{H}$~\citep[see Eq.~(59.5) on p.~173]{LanLif80}. Here $\mu$ is the magnetic permeability of the electron gas. The approximation ${\bf B}\approx {\bf H}$ remains valid with a good accuracy since $\chi\ll 1$. 

Only electrons, which are on the main Landau level $n=0$, contribute to the mean spin, 
\begin{equation}\label{mainLandau}
  \bm{\mathcal{S}}= - n_{0}\hat{\bf n}_\mathrm{B},
  \quad
  \hat{\bf n}_\mathrm{B}=\frac{\bf B}{B},
  \quad
  n_0=\frac{eBp_{\mathrm{F}_e}}{2\pi^2}.
\end{equation}
This fact is valid for a degenerate electron gas, both ultra-relativistic, as in Eq.~(\ref{chiral-limit}), and a non-relativistic, as in Eq.~(\ref{NRplasma}). \citet{Nunokawa:1997dp} obtained that $n_0$ is a part of the total electron density in the degenerate electron gas in Eq.~\eqref{mainLandau},
\begin{equation}\label{totaldensity}
  n_e=\frac{eBp_{\mathrm{F}_e}}{2\pi^2} + \sum_{n=1}^{n_\mathrm{max}}
  \frac{2eB\sqrt{p_{\mathrm{F}_e}^2 - 2eBn}}{2\pi^2},
\end{equation}
where the sum is up to a maximal value $n_\mathrm{max}=[p_{\mathrm{F}_e}^2/(2eB)]$ and $[\dots]$ stays for the integer part of $p_{\mathrm{F}_e}^2/(2eB)$.

The sum in Eq.~(\ref{totaldensity}) is vanishing for strong magnetic fields, $2eB>p_{\mathrm{F}_e}^2$. Thus all electrons populate the main Landau level, $n_e=n_0$. Then, if we consider ultrarelativistic or massless electrons, the total electric current becomes anomalous. It is the manifestation of the CME~\citep{Fuk08,Kha16},
\begin{equation}\label{anomalous_current}
  {\bf j}=\alpha {\bf B}=\frac{\alpha_\mathrm{em}\mu_5}{\pi}{\bf B},\quad\mu_5=(\mu_\mathrm{R} - \mu_\mathrm{L})/2.
\end{equation}
This current in Eq.~\eqref{anomalous_current} is known to produce a forceless magnetic field, ${\bf j}\times {\bf B}=0$. The explicit form of such a field in three dimensions, which is the solution of the equation $\nabla\times {\bf B}=\alpha {\bf B}$, was found by~\citet{Chandra}. 

If we deal with a magnetic field obeying $eB>p_{\mathrm{F}_e}^2/2$, a realistic assumption $p_{\mathrm{F}_e}\sim 100\,{\rm MeV}$ is hard to implement in the core of NS. It results from the fact that the magnetic field, necessary for the condition $n_e=n_0$, is rather strong: $B> 10^4 B_\mathrm{cr}/2\sim 2.2\times 10^{17}\,{\rm G}$. Here $B_\mathrm{cr} = m_e^2/e = 4.4\times10^{13}\,\text{G}$ is the critical magnetic field. 

We can take a moderately strong magnetic field, $m_e^2\ll 2eB\ll p_{\mathrm{F}_e}^2$. This situation can be implemented in magnetars, where $B\sim 10^{15}\,{\rm G}$. Indeed, both electrons and protons in NS have $p_{\mathrm{F}_e}=p_{\mathrm{F}_p}\sim 100\,{\rm MeV}$ in a electroneutral plasma, $n_e=n_p$. In this case, the sum for Landau levels $1\leq n\leq n_\mathrm{max}$ in Eq.~(\ref{totaldensity}) has a greater contribution to the total current compared to the anomalous one. Therefore, the usual transverse components ${\bf B}\perp {\bf j}$ become greater leading to the existence of a Lorentz force. It means that the CME turns out to be negligible since the population of the main Landau level is small\footnote{We shall demonstrate below that such a small correction leads to the very important term for the magnetic helicity dissipation in the situation of the rigid rotation of a magnetar.} in comparison with the the total one~\citep{Nunokawa:1997dp},
\begin{equation}\label{real}
n_e\approx \frac{p_{\mathrm{F}_e}^3}{3\pi^2}\left[1 + \frac{3eB}{2p_{\mathrm{F}_e}^2}\right].
\end{equation}
The correction to the electron number density from the magnetic field in Eq.~\eqref{real} results in the dependence of the chemical potential on the magnetic field; cf. Eq.~\eqref{total}.

\subsection{Approximate description of chiral plasma based on the Berry phase
evolution\label{sec:EVOL}}

Now we study the motion of a chiral fermion in background matter accounting for the electroweak interaction. We shall add the contribution of the electromagnetic field later. One has to consider the Dirac equation for an electron in the form,
\begin{equation}\label{eq:Direqcov}
  \mathrm{i}\dot{\psi}=
  \left[
    (\bm{\alpha}\cdot\hat{\mathbf{p}})+
    V_{\mathrm{R}}^{\mu}\gamma^{0}\gamma_{\mu}P_{\mathrm{R}}+
    V_{\mathrm{L}}^{\mu}\gamma^{0}\gamma_{\mu}P_{\mathrm{L}}
  \right]\psi,
\end{equation}
where $\hat{\mathbf{p}}=-\mathrm{i}\nabla$ is the operator of the electron momentum,
$\psi$ is the wavefunction of an electron,
$\bm{\alpha}=\gamma^{0}\bm{\gamma}$ 
are the Dirac matrices, $P_{\mathrm{L,R}}=(1\pm\gamma^{5})/2$ are
the projectors to the chiral states, and $V_{\mathrm{L},\mathrm{R}}^{\mu}=V_{\mathrm{L},\mathrm{R}}^{\mu}(\mathbf{x})=(V_{\mathrm{L},\mathrm{R}}^{0},\mathbf{V}_{\mathrm{L,R}})$
are the potentials of the interaction of the left and right chiral states
with the inhomogeneous matter.

\citet{DvoSem15} found the explicit expressions for $V_{\mathrm{L},\mathrm{R}}^{0}\sim G_{\mathrm{F}}n_{f}$
for electrons. Here $n_{f}$
is the density of background fermions and $G_{\mathrm{F}}=1.17\times10^{-5}\,\text{GeV}^{-2}$
is the Fermi constant. \citet{DvoStu02} derived that $\mathbf{V}_{\mathrm{L},\mathrm{R}}=V_{\mathrm{L},\mathrm{R}}^{0}\mathbf{v}$ for the arbitrarily moving unpolarized
background matter. Here $\mathbf{v}(\mathbf{x})$ is the velocity of plasma, in which
all background fermions are taken to move as a whole. We can consider the situation when the density
of background matter is also the function of coordinates, $n_{f}=n_{f}(\mathbf{x})$. Since the
electroweak forces violates parity, then $V_{\mathrm{R}}^{\mu}\neq V_{\mathrm{L}}^{\mu}$.

We write down $\psi$ in the form of two chiral projections, $\psi^{\mathrm{T}}=(\psi_{\mathrm{R}},\psi_{\mathrm{L}})$ if we adopt the chiral representation for the Dirac matrices. Using Eq.~(\ref{eq:Direqcov}), one obtains the wave equations for
$\psi_{\mathrm{L,R}}$,
\begin{equation}\label{eq:Direqchirproj}
  \mathrm{i}\dot{\psi}_{\mathrm{L,R}}=
  \left[
    \mp(\bm{\sigma}\cdot\hat{\mathbf{p}})+
    V_{\mathrm{L,R}}^{0}\pm(\bm{\sigma}\cdot\mathbf{V}_{\mathrm{L,R}})
  \right]
  \psi_{\mathrm{L,R}},
\end{equation}
where $\bm{\sigma}$ are the Pauli matrices.

One can hardly find the solution of Eq.~(\ref{eq:Direqchirproj}) if the effective potentials arbitrarily depend on coordinates,
$V_{\mathrm{L},\mathrm{R}}^{\mu}(\mathbf{x})$.
However, we can apply the quasiclassical method, based on the concept of the Berry phase~\citep{Ber84}, for the analysis of the evolution of $\psi_{\mathrm{L,R}}$. For this purpose, we replace the field theory approach with the classical mechanics. Namely, we take
that particles move along certain trajectories. Thus we should choose some canonical coordinates $\{\mathbf{x}(t),\mathbf{p}(t)\}$.
These coordinates can be taken in numerous ways. Therefore, we replace the wave functions $\psi_{\mathrm{L,R}}(\mathbf{x},t)$,
depending on time $t$ and the coordinates $\mathbf{x}$ with effective ones, depending on $t$ only, $\psi_{\mathrm{L,R}}(t)$: $\psi_{\mathrm{L,R}}(\mathbf{x},t)\to\exp(\mathrm{i}\Phi)\psi_{\mathrm{L,R}}(t)$. The dependence on the canonical coordinates is in the phase $\Phi=\Phi(\mathbf{x},\mathbf{p})$, which is not fixed yet.

Basing on Eq.~(\ref{eq:Direqchirproj}), we get that the new
spinors $\psi_{\mathrm{L,R}}=\psi_{\mathrm{L,R}}(t)$ obey the following equation:
\begin{equation}\label{eq:HRL}
  \mathrm{i}\frac{\mathrm{d}}{\mathrm{d}t}{\psi}_{\mathrm{L,R}}=H_{\mathrm{L,R}}\psi_{\mathrm{L,R}},
  \quad
  H_{\mathrm{L,R}}=V_{\mathrm{L,R}}^{0} \mp(\bm{\sigma}\bm{\Pi}),
\end{equation}
where $\bm{\Pi}=\nabla\Phi-\mathbf{V}_{\mathrm{L,R}}$. The Hamiltonians
$H_{\mathrm{L,R}}$ in Eq.~(\ref{eq:HRL}) depend on time only since
we assume that $\mathbf{x}=\mathbf{x}(t)$.

Now we take that $H_{\mathrm{L,R}}$ change adiabatically. In this situation, \citet{Vin90} found that
\begin{equation}\label{eq:psiTheta}
  \psi_{\mathrm{L,R}}(t)=\exp(-\mathrm{i}\Theta_{\mathrm{L,R}})u_{\mathrm{L,R}},
\end{equation}
where $\Theta_{\mathrm{L,R}}=\Theta_{\mathrm{L,R}}(t)$ is the phase introduced by~\citet{Ber84}. In our case, the Berry phase is different for left and right electrons. In Eq.~\eqref{eq:psiTheta}, $u_{\mathrm{L,R}}$ are the constant spinors. Following~\citet{Vin90}, we impose the
constraints on $\psi_{\mathrm{L,R}}$,
\begin{equation}\label{eq:normcond}
  \psi_{\mathrm{L,R}}^{\dagger}\dot{\psi}_{\mathrm{L,R}}=0,
  \quad
  \psi_{\mathrm{L,R}}^{\dagger}\psi_{\mathrm{L,R}}=1,
\end{equation}
which fix their amplitudes and the phases.

\citet{Vin90} suggested to solve the Schr\"odinger
equation,
\begin{equation}\label{eq:Schrequ}
  H_{\mathrm{L,R}}u_{\mathrm{L,R}}(\xi) = E_{\mathrm{L,R}}u_{\mathrm{L,R}}(\xi),
\end{equation}
to find $\Theta_{\mathrm{L,R}}$. In Eq.~\eqref{eq:Schrequ}, the canonical variable $\xi_{a}=\{\mathbf{x},\mathbf{p}\}$,
$a=1,\dots,6$, is taken to be constant: $\xi_{a}=\text{const}$.
The solution of Eq.~\eqref{eq:Schrequ} has the form,
\begin{equation}\label{eq:Energyu}
  E_{\mathrm{L,R}}=|\bm{\Pi}|+V_{\mathrm{L,R}}^{0},
  \quad
  u_{\mathrm{L}}=
  \left(
    \begin{array}{c}
      -e^{-\mathrm{i}\varphi/2}\sin\dfrac{\theta}{2} \\
      e^{\mathrm{i}\varphi/2}\cos\dfrac{\theta}{2}
    \end{array}
  \right),
  \quad
  u_{\mathrm{R}}=
  \left(
    \begin{array}{c}
      e^{-\mathrm{i}\varphi/2}\cos\dfrac{\theta}{2} \\
      e^{\mathrm{i}\varphi/2}\sin\dfrac{\theta}{2}
    \end{array}
  \right),
  \quad
\end{equation}
where the angles $\theta$ and $\varphi$ determine the direction of the vector $\bm{\Pi}$.

Using the results of~\citet{Vin90}, as well as Eqs.~(\ref{eq:HRL})-(\ref{eq:Schrequ}), we obtain that the Berry
phases $\Theta_{\mathrm{L,R}}$ satisfy the equation,
\begin{equation}\label{eq:EqTheta}
  \dot{\Theta}_{\mathrm{L,R}}=-\mathrm{i}u_{\mathrm{L,R}}^{\dagger}
  \frac{\partial u_{\mathrm{L,R}}}{\partial\xi_{a}}\dot{\xi}_{a}.
\end{equation}
\citet{Vin90} found that the solution of Eq.~(\ref{eq:EqTheta}) has the form,
\begin{equation}\label{eq:pathint}
  \Theta_{\mathrm{L,R}}(t)=\int_{C}A_{a}^{(\mathrm{L,R})}(\xi)\mathrm{d}\xi_{a},
\end{equation}
where $C$ is the trajectory of an electron in the phase space and
\begin{equation}\label{eq:Berryconngen}
  \mathrm{i}A_{a}^{(\mathrm{L,R})}=u_{\mathrm{L,R}}^{\dagger}
  \frac{\partial u_{\mathrm{L,R}}}{\partial\xi_{a}},
\end{equation}
is the Berry connection.

Using two three-dimensional vectors, we rewrite the Berry connection in the form, $A_{a}^{(\mathrm{L,R})}=(\bm{\mathcal{B}}_{\mathrm{L,R}},\bm{\mathcal{A}}_{\mathrm{L,R}})$. Basing on Eq.~(\ref{eq:Berryconngen}), we get that
\begin{align}\label{eq:ABRL}
    \mathcal{B}_{i}^{(\mathrm{L,R})} & =
  -\mathrm{i}u_{\mathrm{L,R}}^{\dagger}\frac{\partial u_{\mathrm{L,R}}}{\partial x_{i}}=
  -\mathrm{i}u_{\mathrm{L,R}}^{\dagger}\frac{\partial u_{\mathrm{L,R}}}{\partial\Pi_{j}}
  \frac{\partial\Pi_{j}}{\partial x_{i}},
  \nonumber
  \\
  \mathcal{A}_{i}^{(\mathrm{L,R})} & =
  -\mathrm{i}u_{\mathrm{L,R}}^{\dagger}\frac{\partial u_{\mathrm{L,R}}}{\partial p_{i}}=
  -\mathrm{i}u_{\mathrm{L,R}}^{\dagger}\frac{\partial u_{\mathrm{L,R}}}{\partial\Pi_{j}}  
  \frac{\partial\Pi_{j}}{\partial p_{i}}.
\end{align}
Thus, we replace the integration along the particle trajectory by the time integration. Finally,
we rewrite Eq.~(\ref{eq:pathint}) as
\begin{equation}\label{eq:ThetaAPi}
  \Theta_{\mathrm{L,R}}(t)=\int_{t_{0}}^{t}
  \left(
    \bm{\mathcal{B}}_{\mathrm{L,R}}\dot{\mathbf{x}}+
    \bm{\mathcal{A}}_{\mathrm{L,R}}\dot{\mathbf{p}}
  \right)
  \mathrm{d}t=
  \int_{t_{0}}^{t}\mathcal{A}_{\bm{\Pi}j}^{(\mathrm{L,R})}
  \left[
    \frac{\partial\Pi_{j}}{\partial p_{i}}\dot{p}_{i}+
    \frac{\partial\Pi_{j}}{\partial x_{i}}\dot{x}_{i}
  \right]
  \mathrm{d}t,
\end{equation}
where
\begin{equation}
  \mathrm{i}\mathcal{A}_{\bm{\Pi}j}^{(\mathrm{L,R})} =
  u_{\mathrm{L,R}}^{\dagger}\frac{\partial u_{\mathrm{L,R}}}{\partial\Pi_{j}},
\end{equation}
which stems from Eq.~(\ref{eq:ABRL}).

According to Eq.~(\ref{eq:psiTheta}), the wavefunction evolves as $\psi_{\mathrm{L,R}}(t)\sim\exp(-\mathrm{i}\Theta_{\mathrm{L,R}})$. If we recall that the spinors $u_{\mathrm{L,R}}$
correspond to the energies $E_{\mathrm{L,R}}$ in Eq.~(\ref{eq:Energyu}), as well as accounting for Eq.~(\ref{eq:ThetaAPi}), we can introduce the effective energies,
\begin{equation}\label{eq:Eeff}
  E_{\mathrm{eff}}=\mathcal{A}_{\bm{\Pi}j}^{(\mathrm{L,R})}  
  \left[
    \frac{\partial\Pi_{j}}{\partial p_{i}}\dot{p}_{i}+
    \frac{\partial\Pi_{j}}{\partial x_{i}}\dot{x}_{i}
  \right] +
  E_{\mathrm{L,R}}.
\end{equation}
Now we can take into account the electromagnetic interaction of chiral electrons. We suppose that their electric charge is $e$
and one has the external electromagnetic field with the vector potential $A^{\mu}=(A_{0},\mathbf{A})$.
Basing on Eq.~(\ref{eq:Eeff}), we get the effective actions
for right and left electrons,
\begin{align}\label{eq:SeffRL}
  S_{\mathrm{eff}}^{(\mathrm{L,R})}[\mathbf{x},\mathbf{p}]= &
  \int_{t_{0}}^{t}\mathrm{d}t
  \bigg\{
    \dot{\mathbf{x}}
    \left[
      \mathbf{p}+e\mathbf{A}(\mathbf{x})
    \right]-
    \mathcal{A}_{\bm{\Pi}j}^{(\mathrm{L,R})}
    \left[
      \frac{\partial\Pi_{j}}{\partial p_{i}}\dot{p}_{i}+
      \frac{\partial\Pi_{j}}{\partial x_{i}}\dot{x}_{i}
    \right]
    \notag
    \\
    & -
    \epsilon_{\mathrm{L,R}}(\mathbf{x})-eA_{0}(\mathbf{x})  
  \bigg\},
\end{align}
where, following \citet{SonYam13}, we account for the dependence of the electron
energy on coordinates $\epsilon_{\mathrm{L,R}}(\mathbf{x})$.

We have not specified the canonical coordinates and $\bm{\Pi}$ yet. We take $\{\mathbf{x},\mathbf{p}\}$ to have $\Phi=(\mathbf{px})$. In this situation,
\begin{equation}
  \bm{\Pi}=\mathbf{p}-\mathbf{V}_{\mathrm{L,R}},
  \quad
  \frac{\partial\Pi_{k}}{\partial p_{i}}=\delta_{ik},
  \quad
  \frac{\partial\Pi_{k}}{\partial x_{i}}=-\frac{\partial V_{k}^{(\mathrm{L,R})}}{\partial x_{i}},
\end{equation}
and $\epsilon_{\mathbf{P}}^{(\mathrm{L,R})}(\mathbf{x})=V_{\mathrm{L,R}}^{0}+|\mathbf{P}|+\epsilon_{0}$. Here
where $\epsilon_{0}=\epsilon_{0}(\mathbf{x})$ is due to the coordinates dependence of the electromagnetic field.

The effective action in Eq.~(\ref{eq:SeffRL}) has the form,
\begin{equation}\label{eq:SLRint}
  S_{\mathrm{eff}}^{(\mathrm{L,R})}=\int_{t_{0}}^{t}\mathrm{d}t
  \left\{
    \dot{\mathbf{x}}
    \left[
      \mathbf{p}+e\mathbf{A}(\mathbf{x})
    \right]-
    \bm{\mathcal{A}}_{\mathrm{L,R}}
    \left[
      \dot{\mathbf{p}}-(\dot{\mathbf{x}}\nabla)\mathbf{V}_{\mathrm{L,R}}
    \right]-
    \epsilon_{\mathbf{P}}^{(\mathrm{L,R})}(\mathbf{x})-eA_{0}(\mathbf{x})
  \right\},
\end{equation}
where the index $\mathbf{P}$ in the Berry connection is omitted:
$\bm{\mathcal{A}}_{\mathbf{P}}^{(\mathrm{L,R})}\to\bm{\mathcal{A}}_{\mathrm{L,R}}$.
Basing on the fact that $\dot{\mathbf{P}}=\dot{\mathbf{p}}-(\dot{\mathbf{x}}\nabla)\mathbf{V}_{\mathrm{L,R}}$,
we can recast Eq.~(\ref{eq:SLRint}) as
\begin{equation}\label{eq:SRLfin}
  S_{\mathrm{eff}}^{(\mathrm{L,R})}[\mathbf{x},\mathbf{P}]=
  \int_{t_{0}}^{t}\mathrm{d}t
  \left\{
    \dot{\mathbf{x}}
    \left[
      \mathbf{P}+e\mathbf{A}_{\mathrm{eff}}(\mathbf{x})
    \right]-
    \bm{\mathcal{A}}_{\mathrm{L,R}}\dot{\mathbf{P}}-
    \epsilon_{\mathbf{P}}^{(\mathrm{L,R})}(\mathbf{x})-eA_{0}(\mathbf{x})
  \right\},
\end{equation}
where $\mathbf{A}_{\mathrm{eff}}=\mathbf{A}+\mathbf{V}_{\mathrm{L,R}}/e$. Now we can utilize the new set of the canonical variables $\Xi_{a}=\{\mathbf{x},\mathbf{P}\}$. These canonical variables were used by~\citet{Sil99} to construct kinetic equations
of fermions, electroweakly interacting with inhomogeneous background matter.

%\begin{comment}
%We define the matrix
%\[
%\omega_{ab}=\partial_{a}\omega_{b}-\partial_{b}\omega_{a}=\left(\begin{array}{cc}
%-e\left[\frac{\partial A_{j}^{(eff)}}{\partial x_{i}}-\frac{\partial A_{i}^{(eff)}}{\partial x_{j}}\right] & \delta_{ij}\\
%-\delta_{ij} & \frac{\partial\mathcal{A}_{j}}{\partial P_{i}}-\frac{\partial\mathcal{A}_{i}}{\partial P_{j}}
%\end{array}\right)=\left(\begin{array}{cc}
%-e\varepsilon_{ijk}B_{k}^{(eff)} & \delta_{ij}\\
%-\delta_{ij} & \varepsilon_{ijk}\Omega_{k}
%\end{array}\right).
%\]
%where $\omega_{a}=\left(-\mathbf{P}-e\mathbf{A}_{eff}(x),\bm{\mathcal{A}}_{R,L}\right)$
%and $\partial_{a}=\partial/\partial\Xi_{a}$. The determinant of this
%matrix is $\det(\omega)=(1+e\mathbf{B}_{eff}\bm{\Omega})^{2}$, where
%$\mathbf{B}_{eff}=(\nabla\times\mathbf{A}_{eff})$ and $\bm{\Omega}_{\mathbf{P}}=(\nabla_{\mathbf{P}}\times\bm{\mathcal{A}}_{R,L})$
%is the Berry curvature. The inverse matrix reads
%\[
%\left(\omega^{-1}\right)_{ab}=\frac{1}{1+e\mathbf{B}_{eff}\bm{\Omega}_{\mathbf{P}}}\left(\begin{array}{cc}
%\varepsilon_{ijk}\Omega_{k} & -\delta_{ij}-B_{i}^{(eff)}\Omega_{j}\\
%\delta_{ij}+B_{i}^{(eff)}\Omega_{j} & -e\varepsilon_{ijk}B_{k}^{(eff)}
%\end{array}\right).
%\]
%
%The kinetic equation for the distribution functions $f_{R,L}=f_{R,L}(\Xi_{a},t)$
%has the form,
%\[
%\partial_{t}f_{R,L}-(\omega^{-1})_{ab}\partial_{a}f_{R,L}\partial_{b}H_{eff}^{(R,L)}=0,
%\]
%where $H_{eff}^{(R,L)}=\epsilon_{\mathbf{P}}^{(R,L)}(\mathbf{x})+eA_{0}(\mathbf{x})$.
%It can be rewritten as
%\end{comment}

Applying the approach of~\citet{SonYam13}, as well as using Eq.~(\ref{eq:SRLfin}), we get the kinetic equation for the
distribution functions of left and right fermions, $f_{\mathrm{L,R}}(\mathbf{x},\mathbf{P},t)$, in the form,
\begin{align}\label{eq:kineq}
  \frac{\partial f_{\mathrm{L,R}}}{\partial t}+ &
  \frac{1}{1+e\mathbf{B}_{\mathrm{eff}}\bm{\Omega}}
  \bigg\{
    \left[
      \mathbf{v}+e(\tilde{\mathbf{E}}\times\bm{\Omega}_{\mathbf{P}})+
      e(\mathbf{v}\cdot\bm{\Omega}_{\mathbf{P}})\mathbf{B}_{\mathrm{eff}}
    \right]
    \frac{\partial f_{\mathrm{L,R}}}{\partial\mathbf{x}}
    \nonumber
    \\
    & +
    \left[
      e\tilde{\mathbf{E}}+e(\mathbf{v}\times\mathbf{B}_{\mathrm{eff}})+
      e^{2}(\tilde{\mathbf{E}}\cdot\mathbf{B}_{\mathrm{eff}})\bm{\Omega}_{\mathbf{P}}
    \right]
    \frac{\partial f_{\mathrm{L,R}}}{\partial\mathbf{P}}
  \bigg\} =0,
\end{align}
where $\tilde{\mathbf{E}}=\mathbf{E}_{\mathrm{eff}}-e^{-1}\partial\epsilon_{\mathbf{P}}/\partial\mathbf{x}$,
$\mathbf{v}=\partial\epsilon_{\mathbf{P}}/\partial\mathbf{P}$, $\mathbf{E}_{\mathrm{eff}}=-\nabla A_{0}-\nabla V_{\mathrm{L,R}}^{0}/e$,
$\mathbf{B}_{\mathrm{eff}}=(\nabla\times\mathbf{A}_{\mathrm{eff}})$,
and $\bm{\Omega}_{\mathbf{P}}=(\nabla_{\mathbf{P}}\times\bm{\mathcal{A}}_{\mathrm{L,R}})$
is the Berry curvature. To derive Eq.~(\ref{eq:kineq}) we take $\epsilon_{\mathbf{P}}^{(\mathrm{L,R})}=V_{\mathrm{L,R}}^{0}+\epsilon_{\mathbf{P}}$,
where $\epsilon_{\mathbf{P}}(\mathbf{x})=|\mathbf{P}|+\epsilon_{0}(\mathbf{x})$.

Note that, if $\epsilon_{0}(\mathbf{x})=0$, Eq.~(\ref{eq:kineq})
coincides with that derived by~\citet{DvoSem17,DvoSem18a} for
a neutrino electroweakly interacting with electron-positron plasma
since $\tilde{\mathbf{E}}=\mathbf{E}_{\mathrm{eff}}=\mathbf{E}+\mathbf{E}_{\mathrm{EW}}$
and $\mathbf{B}_{\mathrm{eff}}=\mathbf{B}+\mathbf{B}_{\mathrm{EW}}$,
where $\mathbf{E}_{\mathrm{EW}}=-e^{-1}\nabla V_{\mathrm{L,R}}^{0}$
and $\mathbf{B}_{\mathrm{EW}}=e^{-1}(\nabla\times\mathbf{V}_{\mathrm{L,R}})$
are the effective corrections induced by the electroweak interaction to the
conventional electromagnetic field $F_{\mu\nu}=(\mathbf{E},\mathbf{B})$.

\subsection{Electric current and Adler anomaly in inhomogeneous matter accounting for the electroweak interaction\label{sec:ELCURR}}

In this section, we derive the corrections to the anomalous current
and to the Adler anomaly arising from the electroweak interaction
of chiral fermions with inhomogeneous background matter. We consider
the particular case of the rotating matter.

\citet{SonYam13} defined the number densities
\begin{equation}\label{eq:dens}
  n_{\mathrm{L,R}}=\int\frac{\mathrm{d}^{3}P}{(2\pi)^{3}}
  \left[
    1+e
    \left(
      \mathbf{B}_{\mathrm{eff}}\cdot\bm{\Omega}_{\mathbf{P}}
    \right)
  \right]
  f_{\mathrm{L,R}}.
\end{equation}
and the currents
\begin{align}\label{eq:current}
  \mathbf{j}_{\mathrm{L,R}}= & -\int\frac{\mathrm{d}^{3}P}{(2\pi)^{3}}  
  \bigg[
    \epsilon_{\mathbf{P}}\frac{\partial f_{\mathrm{L,R}}}{\partial\mathbf{P}}+
    e
    \left(
      \bm{\Omega}_{\mathbf{P}}\cdot\frac{\partial f_{\mathrm{L,R}}}{\partial\mathbf{P}}
    \right)
    \epsilon_{\mathbf{P}}\mathbf{B}_{\mathrm{eff}}
    \notag
    \\
    & +
    \epsilon_{\mathbf{P}}
    \left(
      \bm{\Omega}_{\mathbf{P}}\times\frac{\partial f_{\mathrm{L,R}}}{\partial\mathbf{x}}  
    \right)+
    e
    \left(
      \mathbf{E}_{\mathrm{eff}}\times\bm{\Omega}_{\mathbf{P}}
    \right)f_{\mathrm{L,R}}
  \bigg],
\end{align}
\citet{DvoSem17,DvoSem18a} obtained that the quantities in Eqs.~\eqref{eq:dens} and~\eqref{eq:current} satisfy the equation,
\begin{equation}\label{eq:conteq}
  \partial_{t}n_{\mathrm{L,R}}+(\nabla\cdot\mathbf{j}_{\mathrm{L,R}})=
  -e^{2}\int\frac{\mathrm{d}^{3}P}{(2\pi)^{3}}
  \left(
    \bm{\Omega}_{\mathbf{P}}\cdot\frac{\partial f_{\mathrm{L,R}}}{\partial\mathbf{P}}
  \right)
  \left(
    \mathbf{E}_{\mathrm{eff}}\cdot\mathbf{B}_{\mathrm{eff}}
  \right).
\end{equation}
\citet{DvoSem17,DvoSem18a} implied the validity of the Maxwell equations for the effective electromagnetic fields, $(\nabla\cdot\mathbf{B}_{\mathrm{eff}})=0$ and $\partial_{t}\mathbf{B}_{\mathrm{eff}}+(\nabla\times\mathbf{E}_{\mathrm{eff}})=0$ to derive Eq.~(\ref{eq:normcond}).

We shall consider the homogeneous distribution functions: $\partial f_{\mathrm{L,R}}/\partial\mathbf{x}=0$.
In this situation, $\epsilon_{\mathbf{P}}=\mu_{\mathrm{L,R}}+\dotsb$,
where $\mu_{\mathrm{L,R}}$ are the chemical potentials of right and
left electrons. Moreover the zero order terms in the Fermi constant
$G_{\mathrm{F}}$ are kept. One gets that the anomalous part of the
current $\mathbf{J}_{\mathrm{L,R}}=e\mathbf{j}_{\mathrm{L,R}}$ is expressed as
\begin{align}\label{eq:JRL}
  \mathbf{J}_{\mathrm{L,R}}= &
  -\mu_{\mathrm{L,R}}e^{2}\mathbf{B}_{\mathrm{eff}}
  \int\frac{\mathrm{d}^{3}P}{(2\pi)^{3}}
  \left(
    \bm{\Omega}_{\mathbf{P}}\cdot\frac{\partial f_{\mathrm{L,R}}}{\partial\mathbf{P}}
  \right) =
  \mp\frac{e^{2}}{4\pi^{2}}\mu_{\mathrm{L,R}}\mathbf{B}_{\mathrm{eff}}
  \notag
  \\
  & =
  \mp\frac{e^{2}}{4\pi^{2}}\mu_{\mathrm{L,R}}
  \left[
    \mathbf{B}+\frac{1}{e}(\nabla\times\mathbf{V}_{\mathrm{L,R}})
  \right].
\end{align}
The first term in Eq.~(\ref{eq:JRL}), namely $\mathbf{J}_{\mathrm{L,R}}\propto\mu_{\mathrm{L,R}}\mathbf{B}$,
reproduces the CME~\citep{Fuk08}. We take the following distribution function:

\begin{equation}\label{eq:distrfun}
  f_{\mathrm{L,R}}(\mathbf{P})=
  \frac{1}{
    \exp[\beta(|\mathbf{P}|-\mu_{\mathrm{L,R}})]+1
  },
  \quad
  \frac{\partial f_{\mathrm{L,R}}}{\partial\mathbf{P}}=
  -\frac{\mathbf{P}}{|\mathbf{P}|}\frac{\beta\exp[\beta(|\mathbf{P}|-\mu_{\mathrm{L,R}})]}{
  \left\{
    \exp[\beta(|\mathbf{P}|-\mu_{\mathrm{L,R}})]+1
  \right\}^{2}
  },
\end{equation}
to derive Eq.~(\ref{eq:JRL}). Here $\beta=1/T$ is the inverted temperature. We consider the degenerate plasma with
$\beta\to\infty$ and $\nabla_{\mathbf{P}}f_{\mathrm{L,R}}\to\delta(|\mathbf{P}|-\mu_{\mathrm{L,R}})$ in Eqs.~(\ref{eq:JRL})
and~(\ref{eq:distrfun}). Moreover we take into account the result of~\citet{DvoSem17} that $\bm{\Omega}_{\mathbf{P}}=\mp\mathbf{P}/2|\mathbf{P}|^{3}$.

The correction to the Adler anomaly can be calculated analogously Eq.~(\ref{eq:JRL}) using Eq.~(\ref{eq:conteq}). First, one supposes
that the density of background fermions is homogeneous, i.e.
$\nabla V_{\mathrm{L,R}}^{0}=0$. In this situation, $\mathbf{E}_{\mathrm{eff}}=\mathbf{E}$,
and
\begin{equation}
  \partial_{t}n_{\mathrm{L,R}}+(\nabla\cdot\mathbf{j}_{\mathrm{L,R}})=
  \mp\frac{e^{2}}{4\pi^{2}}
  \left(
    \mathbf{E}\cdot\mathbf{B}_{\mathrm{eff}}
  \right)=
  \mp\frac{e^{2}}{4\pi^{2}}
  \left[
    \left(
      \mathbf{E}\cdot\mathbf{B}
    \right)+
    \frac{1}{e}
    \left(
      \mathbf{E}\cdot\nabla\times\mathbf{V}_{\mathrm{L,R}}
    \right)
  \right].
\end{equation}
Integrating this expression and neglecting the
boundary effects, we get that
\begin{equation}\label{eq:n5gen}
  \frac{\mathrm{d}n_{5}}{\mathrm{d}t}=\frac{e^{2}}{2\pi^{2}V}\int
  \left[
    \left(
      \mathbf{E}\cdot\mathbf{B}
    \right)-
    \frac{1}{e}
    \left(
      \mathbf{E}\cdot\nabla\times\mathbf{V}_{5}
    \right)
  \right]
  \mathrm{d}^{3}x -
  \Gamma n_5,
\end{equation}
where $n_{5}=n_{\mathrm{R}}-n_{\mathrm{L}}$ is the imbalance of chiral electrons and
$\mathbf{V}_{5}=(\mathbf{V}_{\mathrm{L}}-\mathbf{V}_{\mathrm{R}})/2$. Moreover, one has that $(\nabla\times \mathbf{V}_{5}) = V_{5}\bm{\omega}$,
where $V_{5}=(V_{\mathrm{L}}^{0}-V_{\mathrm{R}}^{0})/2$. Here $\bm{\omega}$
is the angular velocity.

We take into account the fact that electrons are ultrarelativistic but not massless particles in Eq.~\eqref{eq:n5gen}. It means that their helicity is changed in collisions with other fermions. Thus we introduce the term $-\Gamma n_5$ in the rhs of Eq.~\eqref{eq:n5gen}. \citet{Dvo16b} computed the spin-flip rate $\Gamma\sim 10^{11}\,\text{s}^{-1}$ in degenerate matter of old NSs.

The correction to the anomalous current owing to the electroweak interaction has the form, $\delta\mathbf{J}_{\mathrm{anom}}=(\mathbf{J}_{\mathrm{L}}+\mathbf{J}_{\mathrm{R}})\propto\nabla\times(\mathbf{V}_{\mathrm{R}}-\mathbf{V}_{\mathrm{L}})\sim V_{5}\bm{\omega}$.
This expression in Eq.~(\ref{eq:JRL}) reproduces the result of~\citet{Dvo15}
except the factor $1/2\pi$ missed  there. \citet{Dvo15} named the generation
of such anomalous current as the galvano-rotational effect.
%\begin{comment}
%However, it should be considered as the electroweak correction to
%the chiral vortical effect (the CVE): $\mathbf{J}\parallel\bm{\omega}$.
%We mention that corrections to the CVE were claimed in Ref.~\cite{Vil80}
%to be vanishing.
%\end{comment}

\section{Astrophysical applications}

In this section, we apply the results of Sec.~\ref{sec:THBACKGR} to study the evolution of magnetic fields in compact stars. In Secs.~\ref{sec:QUANTCORR}-\ref{sec:HELEVOL}, we analyze the magnetic helicity evolution in NS taking into account the polarization of relativistic electrons. In Sec.~\ref{sec:APPL}, we study the flux of the magnetic helicity through the equator of a rotating NS and propose the possible explanation of the observed magnetic cycles of some pulsars.

\subsection{The quantum surface correction to the chiral anomaly in finite volume of a neutron star\label{sec:QUANTCORR}}

In this section, we analyze the system of AMHD equations for chiral fermions. Then we study the modification of these equations when one accounts for a particle mass. In particular, we consider the contribution of the nonzero fermion mass to the Adler anomaly.

\citet{PavLeiSig17} formulated the full set of AMHD equations,
\begin{align}
  \rho
  \left[
    \frac{\partial \mathbf{v}}{\partial t} + (\mathbf{v} \nabla) \mathbf{v} -
    \nu \nabla^2 \mathbf{v}
  \right] = & 
  - \nabla p + \sigma_\mathrm{cond} [(\mathbf{E} \times \mathbf{B}) +
  (\mathbf{v} \times \mathbf{B}) \times \mathbf{B}],
  \notag
  \\
  \frac{\partial \rho}{\partial t} + \nabla (\rho \mathbf{v}) = & 0,
  \label{eq:hyd}
  \\
  \frac{\partial \mathbf{B}}{\partial t} = & - (\nabla \times \mathbf{E}),
  \label{eq:ind}
  \\
  (\nabla \times \mathbf{B}) = & \sigma_\mathrm{cond} 
  \left[
    \mathbf{E} - \frac{e^2}{2\pi^2\sigma_\mathrm{cond}} \mu_5 \mathbf{B} +
    (\mathbf{v} \times \mathbf{B})
  \right],
  \label{eq:CME}
  \\
  \frac{\mathrm{d} \mu_5}{\mathrm{d}t} = &
  - \frac{e^2}{4\mu_e^2} \frac{\mathrm{d} h}{\mathrm{d}t},
  \label{eq:triangle}
\end{align}
where $\rho$ is the density of matter, $\mathbf{v}$ is the velocity of plasma, $p$ is the pressure, $\sigma_\mathrm{cond}$ is the electric conductivity, $\mu_5=(\mu_{e\mathrm{R}} - \mu_{e\mathrm{L}})/2$ is the imbalance between chiral right and left electrons, $h$ is the density of the magnetic helicity, $\mu_e$ is the mean chemical potential of the electron gas, and $\nu$ is the coefficient of the viscosity.

Continuity and Navier-Stokes equations are given in Eq.~\eqref{eq:hyd}. In Eq.~\eqref{eq:CME}, the anomalous current due to the CME is taken into account in the Maxwell equation. The displacement current is omitted there. The induction equation, coming from Eqs.~\eqref{eq:ind} and~\eqref{eq:CME}, is changed due to the CME,
\begin{equation}\label{Faraday}
  \frac{\partial \mathbf{B}}{\partial t} =
  \nabla\times (\mathbf{v}\times \mathbf{B}) +
  \frac{1}{\sigma_\mathrm{cond}}\nabla^2\mathbf{B} -
  \frac{e^2}{2\pi^2\sigma_\mathrm{cond}}\nabla\times (\mu_5\mathbf{B}).
\end{equation}
Equation~\eqref{eq:triangle} results from the Adler-Bell-Jackiw anomaly, which is discussed below, for massless particles. Note that we suppose that the electron gas is degenerate in Eq.~\eqref{eq:triangle}. We will change Eq.~\eqref{eq:triangle} to take into account a nonzero mass of electrons.

We shall examine which terms are important for the Adler anomaly in a finite volume, for instance, in NS.
In particular, we compute the correction due to the averaged mass term $2\mathrm{i}m_e\langle \bar{\psi}\gamma_5\psi\rangle$ to the Adler-Bell-Jackiw anomaly in QED known as the nonconservation of the axial current,
\begin{equation}\label{pseudovector}
\frac{\partial }{\partial x^{\mu}}\bar{\psi}\gamma^{\mu}\gamma^5\psi=\frac{\partial j_\mathrm{R}^{\mu}}{\partial x^{\mu}} - \frac{\partial j_\mathrm{L}^{\mu}}{\partial x^{\mu}} = 2\mathrm{i}m_e\bar{\psi}\gamma_5\psi + \frac{2e^2}{16\pi^2}F_{\mu\nu}\tilde{F}^{\mu\nu},
\end{equation}
where left and right currents are not conserved separately even for massless fermions due to the triangle anomaly in the presence of an elecromagnetic field,
\begin{equation}\label{rightleft}
\frac{\partial j_\mathrm{L,R}^{\mu}}{\partial x^{\mu}}=\mp \mathrm{i}m_e\bar{\psi}\gamma_5\psi \mp \frac{e^2}{16\pi^2}F_{\mu\nu}\tilde{F}^{\mu\nu}.
\end{equation}
Using the fact that $\gamma^0\bm{\gamma}\gamma^5 = \bm{\Sigma}$ and averaging the lhs in Eq.~(\ref{pseudovector}), one obtains
\begin{equation}\label{lhs}
  \int\frac{\mathrm{d}^3x}{V}\frac{\partial }{\partial x^{\mu}}
  \langle\bar{\psi}\gamma^{\mu}\gamma^5\psi\rangle =
  \frac{{\rm d}}{{\rm d}t}(n_\mathrm{R} - n_\mathrm{L}) +
  \frac{1}{V}\oint \mathrm{d}^2S(\bm{\mathcal{S}}\cdot{\bf n}),
\end{equation}
where $\bm{\mathcal{S}}$ is the averaged spin given in Eq.~(\ref{chiral-limit}) for massless fermions.
Averaging the rhs of Eq.~(\ref{pseudovector}), we get
\begin{equation}\label{rhs}
  \dots = -\frac{1}{V}\oint_S (\bm{\mathcal{S}}_{5}\cdot {\bf n}) \mathrm{d}^2S +
  \frac{2\alpha_\mathrm{em}}{\pi}\int \frac{\mathrm{d}^3x}{V}({\bf E}\cdot{\bf B}),
\end{equation}
where the first term $\sim \bm{\mathcal{S}}_5$ is determined by the averaged pseudoscalar,
\begin{equation}\label{mean5}
2\mathrm{i}m_e\int \frac{\mathrm{d}^3x}{V}\langle\bar{\psi}\gamma_5\psi\rangle=
-\int\frac{\mathrm{d}^3x}{V}(\nabla\cdot\bm{\mathcal{S}}_{5}({\bf x},t)).
\end{equation}
The second term in Eq.~\eqref{rhs} gives the dissipation of the magnetic helicity according to the standard MHD~\citep{Pri16},
\begin{equation}\label{standard1}
 2\int \frac{\mathrm{d}^3x}{V}({\bf E}\cdot{\bf B})= -\frac{{\rm d}h}{{\rm d}t} - \oint({\bf n}\cdot\left[{\bf B}A_0 +{\bf E}\times {\bf A}\right])\frac{\mathrm{d}^2S}{V}.
\end{equation}
Equation~(\ref{mean5}) results from the spin distribution functions in a weakly inhomogeneous electron-positron plasma:
\begin{equation}\label{magnetization}
\bm{\mathcal{S}}_{5}({\bf x},t)=-\int\frac{\mathrm{d}^3p}{\gamma(2\pi)^3}\left({\bf S}^{(e)}({\bf p},{\bf x},t) - {\bf S}^{(\bar{e})}({\bf p},{\bf x},t)\right)\left(\frac{2}{3} + \frac{1}{3\gamma}\right).
\end{equation}
\citet{DvoSem18b} provided the detailed derivation of Eq.~\eqref{magnetization}.

\citet{Silin} derived the equilibrium part of the total Wigner spin distribution function of electrons, resulting from the paramagnetic contribution in a inhomogeneous magnetic field,
\begin{equation}\label{spin_0}
  {\bf S}^{(e)}_\mathrm{eq}(\varepsilon_p,{\bf x},t)=
  \frac{\mu_\mathrm{B} {\bf H}({\bf x},t)}{\gamma}
  \frac{{\rm d}f^{(e)}_\mathrm{eq}(\varepsilon_p)}{{\rm d}\varepsilon_p},
  \quad
  \varepsilon_p=\sqrt{p^2 + m_e^2}=\gamma m_e,
\end{equation}
which are used in Eq.~\eqref{magnetization}: ${\bf S}^{(e)}({\bf p},{\bf x},t)={\bf S}^{(e)}_\mathrm{eq}(\varepsilon_p,{\bf x},t) + \delta {\bf S}^{(e)}({\bf p},{\bf x},t)$.

Basing on Eqs.~(\ref{lhs}) and~(\ref{rhs}), as well as taking into account Eq.~(\ref{standard1}) for the magnetic helicity density $h(t)=\smallint \tfrac{\mathrm{d}^3x}{V}({\bf A}\cdot{\bf B})$, we obtain the master equation,
\begin{align}\label{new_law}
 \frac{{\rm d}}{{\rm d}t}
  \left(
    n_\mathrm{R} - n_\mathrm{L} + \frac{\alpha_\mathrm{em}}{\pi}h
  \right) =
  - \frac{\alpha_\mathrm{em}}{\pi V}\oint_S([{\bf E}\times {\bf A} + A_0{\bf B}]\cdot{\bf n})\mathrm{d}^2S - 
  \oint_S
  ([ \bm{\mathcal{S}} + \bm{\mathcal{S}}_{5}]\cdot {\bf n})
  \frac{\mathrm{d}^2S}{V}.
\end{align}
If one neglects the surface terms in a plasma of chiral massless particles, we obtain the conservation law for the sum of the chiral imbalance, $n_\mathrm{R} - n_\mathrm{L}$, and the density of the magnetic helicity $h$, accounting for the factor $\alpha_\mathrm{em}/\pi$,
\begin{equation}\label{CME}
\frac{{\rm d}}{{\rm d}t}\left(n_\mathrm{R} - n_\mathrm{L} +\frac{\alpha_\mathrm{em}}{\pi}h\right)=0.
\end{equation}
Equation~(\ref{CME}) is crucial for AMHD since it describes the evolution of the chiral imbalance, $\sim {\rm d}\mu_5/{\rm d}t$, in a magnetized plasma~\citep{Fuk08,BoyFroRuc12}: a change of the imbalance
%(for instance, due to the spin-flip)
results in the production of the magnetic helicity, and vice versa. 

Equation~(\ref{new_law}) includes the surface term known in classical MHD. It contains the electromagnetic fields only. The new quantum correction results from the sum of the spin terms, $\bm{\mathcal{S}}_\mathrm{eff}=\bm{\mathcal{S}} + \bm{\mathcal{S}}_5$. Below, we shall examine the latter term.

It is worth to be mentioned that the equilibrium distribution function of spins for the positron gas possesses the same positive sign as that for electrons in Eq.~(\ref{spin_0}),
\begin{equation}\label{spin_00}
{\bf S}^{(\bar{e})}_\mathrm{eq}(\varepsilon_p,{\bf x},t)= \frac{\mu_\mathrm{B} {\bf H}({\bf x},t)}{\gamma}\frac{{\rm d}f^{(\bar{e})}_\mathrm{eq}(\varepsilon_p)}{{\rm d}\varepsilon_p}.
\end{equation}
Thus the pseudovector in Eq.~(\ref{magnetization}) is given by the difference of the particle and antiparticle contributions, due to the permutation of the operators $\hat{d}\hat{d}^+\to - \hat{d}^+\hat{d}$ while deriving Eq.~(\ref{magnetization}),
\begin{equation}\label{magnetization2}
  \bm{\mathcal{S}}_{5}({\bf x},t)= - \frac{\mu_\mathrm{B} m_e^2{\bf H}({\bf x},t)}{2\pi^2}\int \frac{p^2\mathrm{d}p}{\varepsilon_p^2}\left(\frac{{\rm d}f^{(e)}_\mathrm{eq}}{{\rm d}\varepsilon_p} - \frac{{\rm d}f^{(\bar{e})}_\mathrm{eq}}{{\rm d}\varepsilon_p}\right)\left(\frac{2}{3} + \frac{1}{3\gamma}\right).
\end{equation}
If we integrate by parts in Eq.~(\ref{magnetization2}) and separate the Lande factor $g_s=2$ from the Fermi distributions functions $f^{(e,\bar{e})}(\varepsilon_p)=g_s[\exp (\varepsilon_p \mp \mu_e)/T +1]^{-1}$, we can rewrite the effective averaged spin $\bm{\mathcal{S}}_\mathrm{eff}=\bm{\mathcal{S}} + \bm{\mathcal{S}}_5$ contributing the surface term in Eq.~(\ref{new_law}), as
\begin{align}\label{quantum_correction}
  \bm{\mathcal{S}}_\mathrm{eff} = &
  - \frac{e{\bf H}}{2\pi^2}\int_0^{\infty}\mathrm{d}p
  \left[
    1 -
    \frac{1}{3\gamma^2}
    \left(
      \frac{2}{\gamma} + \frac{2}{\gamma^2} - 1
    \right)
  \right]
  \notag
  \\
  & \times
  \left(
    \frac{1}{\exp [(\varepsilon_p - \mu_e)/T] +1} -
    \frac{1}{\exp [(\varepsilon_p + \mu_e)/T] +1}
  \right).
\end{align}
One has that $\varepsilon_p=\gamma m_e\approx m_e$ in a nonrelativistic plasma. Thus, the total spin effect is vanishing, $\bm{\mathcal{S}}_\mathrm{eff}=\bm{\mathcal{S}} + \bm{\mathcal{S}}_5  \to 0$, since both spin terms cancel each other.  For instance, one gets that
$\bm{\mathcal{S}} = - \bm{\mathcal{S}}_5 = - 2 e m_e v_{\mathrm{F}_e}{\bf B}/4\pi^2$, where $v_{\mathrm{F}_e}=p_{\mathrm{F}_e}/m_e\ll 1$ in a degenerate electron gas, where the positron contribution is absent. The first correction to the last equality appears owing to the decomposition $\gamma^{-1}\approx 1 - v^2/2$. It is quite small, $\bm{\mathcal{S}}_\mathrm{eff} = e m_e v_{\mathrm{F}_e}^3{\bf B}/3\pi^2$. This correction is valid for weak magnetic fields, $m_e^2\gg p_{\mathrm{F}_e}^2\gg eB$, where the Wentzel-Kramers-Brillouin (WKB) approximation with large Landau numbers, $n\gg 1$, used for the calculation of $\bm{\mathcal{S}}_5$, remains correct.

In the case of ultrarelativistic chiral plasma with $\gamma\gg 1$, one gets from Eq.~(\ref{quantum_correction}) that $\bm{\mathcal{S}}_\mathrm{eff} \approx \bm{\mathcal{S}}= - e\mu_e(r,\theta) {\bf B}/2\pi^2$; cf. Eq.~(\ref{chiral-limit}). Hence the effect of the magnetization becomes great under the condition $\mu_e( r,\theta)\gg m_e$, which can be implemented, for instance, in the NS core. It should be noted that, here, we use an inhomogeneous chemical potential $\mu_e(r,\theta)$  that corresponds to a realistic profile of spherically symmetric electron density in NS~\citep{Lattimer}, $n_e(r)=n_\mathrm{core}Y_e[1 - r^2/R_\mathrm{NS}^2]$, where $Y_e=0.04$ is the electron abundance and $n_\mathrm{core}\simeq 10^{38}\,\text{cm}^{-3}$ is the density in the center of a star.

For example, the anisotropy for the chemical potential results from the magnetic field correction proportional to the total magnetic field entering Landau levels,\footnote{We take a slightly nonuniform magnetic field, i.e., it can be uniform at microscopic scales that are smaller than the mean distance between particles in matter, $L< (n_e)^{-1/3}$. This requirement is needed to receive the energy levels in Eq.~\eqref{levels}.
%as it should be for Landau's task in quantum mechanics,
Nevertheless ${\bf B}$ is taken to be nonuniform at macroscopic length scales similar to $R_\mathrm{NS}$.} $ B(r,\theta)=\sqrt{B_{p}^2(r,\theta) + B_{t}^2(r,\theta)}$, in rotating a star, which is axially symmetric. Here $B_{t,p}$ are the toroidal and poloidal components of the magnetic field (see Sec.~\ref{sec:HELEVOL}). This anisotropy is derived from inverted Eq.~(\ref{real}),\footnote{One means the cubic Eq.~(\ref{real}) written down as $\mu^3_e + 3eB\mu_e/2 - 3\pi^2n_e=0$, where $\mu_e\equiv p_{\mathrm{F}_e}$. The parameter $D=q^2 + p^3= (eB)^3/8 + 9\pi^4n_e^2/4>0$ is positive for the Cartan solution. Thus, we have one real root of this equation given in Eq.~(\ref{anisotrop}), as well as two complex self adjoint ones.}
\begin{equation}\label{anisotrop}
  \mu_e(r,\theta)=[3\pi^2n_e(r)]^{1/3}\left[1 - \frac{eB(r,\theta)}{2(3\pi^2n_e(r))^{2/3}}\right].
\end{equation}
Otherwise,  the surface integral
\begin{equation}\label{profit}
  \oint_S(\bm{\mathcal{S}}\cdot{\bf n})\frac{\mathrm{d}^2S}{V}=
  - \frac{e}{2\pi^2 V}
  \oint_S\mu_e(r,\theta)({\bf B}\cdot{\bf n})\mathrm{d}^2S,
\end{equation}
remaining in Eq.~(\ref{new_law}) for NS, is vanishing for the uniform spherically symmetric $\mu_e(r)_{r=R}=\text{const}$ because of the Gauss law, $\int \mathrm{d}^3x (\nabla\cdot {\bf B})=\oint ({\bf B}\cdot{\bf n})\mathrm{d}^2S=0$.

\subsection{Magnetic helicity evolution as the averaged spin flux through the boundary of a domain\label{sec:HELDISS}}

We consider a realistic situation when the chiral imbalance is vanishing, $n_\mathrm{R} - n_\mathrm{L}\to 0$, in relativistic plasmas taking into account the nonzero electron mass, $m_e\neq 0$, because of the spin-flip. It is necessary to reconcile the evolution of the magnetic helicity density in Eq.~(\ref{standard1}) and statistically averaged Adler anomaly in chiral medium given in Eq.~(\ref{new_law}). \citet{DvoSem15,Dvo16b} showed that this situation happens very soon in the core of a young NS, during $\sim 10^{-12}\,\text{s}$. It takes place even when an initial difference is positive $n_\mathrm{R} (t_0) - n_\mathrm{L}(t_0) > 0$, which can arise in a protoneutron star due to the direct Urca-process, $p+ e_\mathrm{L}^-\to n + \nu_{e\mathrm{L}}$. \citet{BoyFroRuc12} found the similar decay of the chiral imbalance $2\mu_5=\mu_\mathrm{R} - \mu_\mathrm{L}\to 0$ owing to the spin-flip in the cooling universe with the temperatures below $T=10\,{\rm MeV}$. 

Then, at time $t\gg t_0$, we should modify the standard MHD Eq.~(\ref{standard1}) due to Eq.~(\ref{new_law}), accounting for
the magnetic flux through the volume surface weighted by the nonuniform chemical potential $\mu_e(r,\theta)$ in Eq.~(\ref{anisotrop}) that enters the quantum (magnetization) term  in Eq.~(\ref{profit}),
\begin{align}\label{dissipation}
  \oint  \mathrm{d}^2S
  \left(
    \bm{\mathcal{S}}_\mathrm{eff}\cdot{\bf n}
  \right)
  \approx
  & 
  - \frac{e}{2\pi^2}
  \oint_S\mu_e(r,\theta)({\bf B}\cdot {\bf n})\mathrm{d}^2S
  \nonumber
  \\
  & =
  \frac{\alpha_\mathrm{em}}{\pi [3\pi^2n_e(R)]^{1/3}}
  \oint B(R,\theta)({\bf B}\cdot {\bf n})\mathrm{d}^2S.
\end{align}
Here the nonuniform electron density $n_e (R)=n_\mathrm{core}Y_e(1 - R^2/R^2_\text{NS})$ should be large enough to obey the inequality $2eB(R,\theta)\ll [3\pi^2n_e (R)]^{2/3}$ at the surface with radius $R< R_\mathrm{NS}$ since the decomposition is made over the small parameter $2eB/\mu_e^2\ll 1$ in Eq.~(\ref{real}).
Then, we can generalize standard Eq.~(\ref{standard1}) due to the  Adler-Bell-Jackiw anomaly accounting for the additional quantum correction in Eq.~(\ref{dissipation}),
\begin{equation}\label{standard3}
\frac{{\rm d}H}{{\rm d}t}= - 2\int_V\mathrm{d}^3x ({\bf E}\cdot{\bf B})-\oint_S
({\bf n}\cdot[A_0{\bf B} + {\bf E}\times {\bf A}])\mathrm{d}^2S + \frac{1}{\mu_e(R)}\oint_S B(R,\theta)B_r(R,\theta)\mathrm{d}^2S,
\end{equation}
where $B(R,\theta)=\sqrt{B_r^2 + B_{\theta}^2 + B_{\varphi}^2}$ is the total magnetic field strength entering Landau levels. We also use a short notation $\mu_e(R)=[3\pi^2n_e(R)]^{1/3}$.

\citet{Pri16} claims that the volume term in the lhs in Eq.~(\ref{standard1}) is negligible at the evolution times smaller than the diffusion one: $t\ll \tau_\mathrm{D}\sim L^2\sigma_\mathrm{cond}$. Thus the behavior of the magnetic helicity is driven only by the second term in Eq.~(\ref{standard3}) or the last term in Eq.~(\ref{standard1}). \citet{SchSht18} obtained that the diffusion time is great, $\tau_\mathrm{D}\simeq 30\,\text{yr}(L/\text{cm})^2$, owing to a big electric conductivity in NS, $\sigma_\mathrm{cond} \sim 10^9\,{\rm MeV}$, for $T = 10^8\,\text{K}$. For example, the diffusion time is longer than the age of the universe, $\tau_\mathrm{D} \approx 3 \times 10^{13}\,\text{yr}\gg t_\mathrm{Univ}=1.4\times 10^{10}\,\text{yr}$, for the maximal length scale $L=R_\mathrm{NS}=10^6\,\text{cm}$. Hence for such large scales there is no a reason to take into account both the magnetic helicity diffusion and the quantum term in Eq.~(\ref{dissipation}) which both could be essential only at times $t> \tau_\mathrm{D}$. However, at small scales $L\ll R_\mathrm{NS}$, the magnetic helicity diffusion time is less than the age of young magnetars $\sim 10^3\,\text{yr}$, e.g., for $L=1\,\text{cm}$, $\tau_\mathrm{D}\sim 30\,\text{yr}\ll t\sim 10^3\,\text{yr}$. Therefore the quantum contribution to the evolution Eq.~(\ref{standard3}), missed in classical approach~\citep{Pri16}, can be essential for small-scale magnetic fields in NS at times $t>\tau_\mathrm{D}$. 

\subsection{Evolution of the magnetic helicity in NS\label{sec:HELEVOL}}
 
The study of the magnetic helicity evolution is important for a possible  reconnection of magnetic field lines near the NS surface  happening mostly outside the crust in the NS magnetosphere. This process, in its turn, could explain gamma or X-ray flares reported by~\citet{KasBel17} to take place in magnetars. However, it  is interesting to study also how dissipation of the magnetic helicity proceeds inside NS below the crust.

Using the gauge $A_0=0$ and $(\nabla\cdot{\bf A})=0$ in Eq.~(\ref{standard1}), valid in classical MHD, and substituting the Ohm law ${\bf E}= - {\bf v}\times{\bf B} + {\bf j}/\sigma_\mathrm{cond}$,  one can rederive the result of~\citet{Pri16},
\begin{equation}\label{eq:helevolclass}
\frac{{\rm d}H}{{\rm d}t}= - 2\sigma_\mathrm{cond}^{-1}\int {\bf j}\cdot{\bf B}\mathrm{d}^3x -\oint_S[({\bf B}\cdot{\bf A})({\bf v}\cdot{\bf n}) - ({\bf v}\cdot{\bf A})({\bf B}\cdot{\bf n})]\mathrm{d}^2S, 
\end{equation}
which is generalized in Eq.~(\ref{standard3}) in this gauge,
\begin{eqnarray}\label{surface_new}
  \frac{{\rm d}H}{{\rm d}t}= &&- 2\sigma_\mathrm{cond}^{-1}\int {\bf j}\cdot{\bf B}\mathrm{d}^3x
  -
  \oint_S[({\bf B}\cdot{\bf A})({\bf v}\cdot{\bf n}) -
  ({\bf v}\cdot{\bf A})({\bf B}\cdot{\bf n})]\mathrm{d}^2S \nonumber\\&&+ 
    \frac{1}{\mu_e(R)}
  \oint_S B(R,\theta)({\bf B}\cdot {\bf n})\mathrm{d}^2S .
\end{eqnarray}
The last term in the rhs of Eq.~\eqref{surface_new} results from the quantum contribution of the averaged spin flux through the surface $R_\text{NS}>R$. We shall show below that it should be essential at small scales. Notice that the magnetic helicity in Eq.~(\ref{surface_new}) is conserved, ${\rm d}H/{\rm d}t=0$, inside a closed volume, where $({\bf v}\cdot{\bf n})=0$ and $({\bf B}\cdot {\bf n})=0$. It happens when one neglects the volume losses for an ideal plasma, $\sigma_{\rm cond}\to \infty$. Both quantum and classical surface terms are different from zero if a magnetic field penetrates the domain volume, i.e. when $({\bf B}\cdot {\bf n})\neq 0$.

\begin{figure}
  \centering
  \includegraphics[scale=0.5]{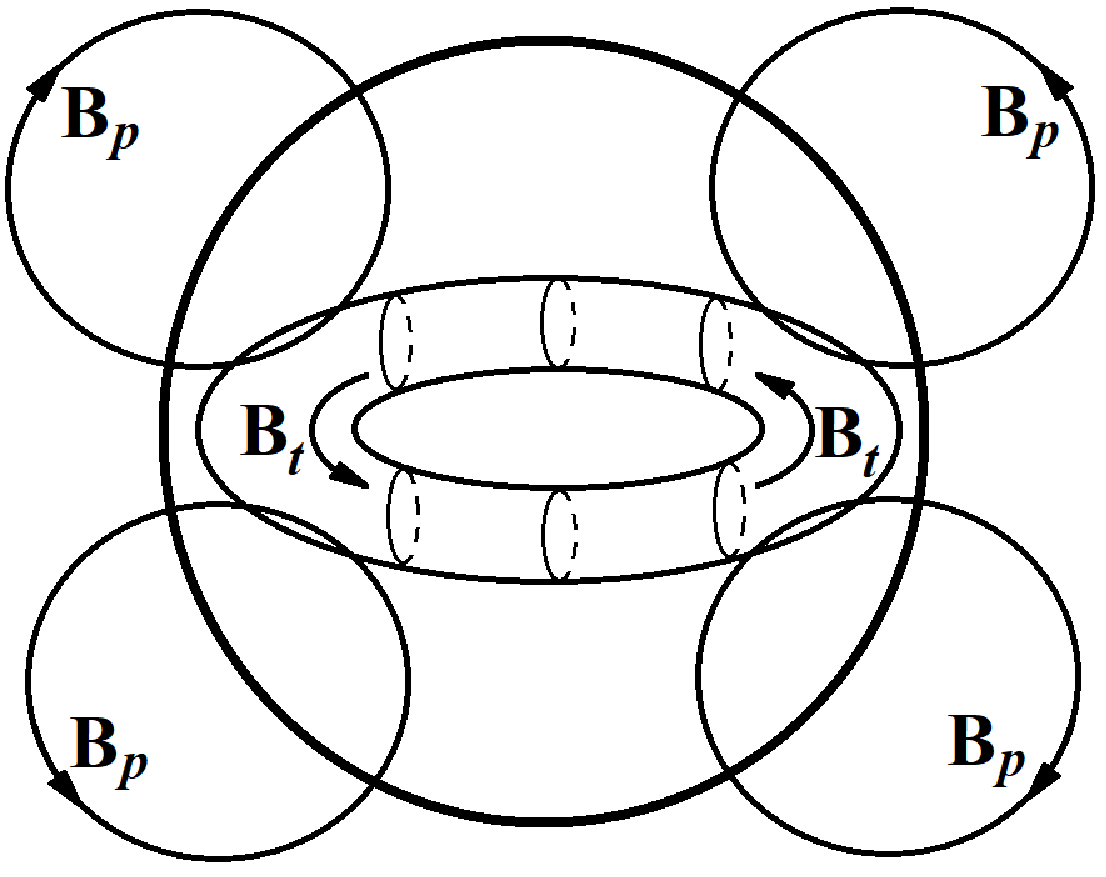}
  \caption{The structure of the magnetic field presented
  in Eq.~\eqref{model}. The  poloidal component of the magnetic field reads
  $\mathbf{B}_p = B_{p}(r)[\cos 2\theta {\bf e}_r + \sin 2\theta {\bf e}_{\theta}]$. The toroidal one is
  $\mathbf{B}_t = B_{\varphi}(r)\cos \theta {\bf e}_{\varphi}$.  
  The figure is taken from the work by~\citet{DvoSem18b}.
  \label{fig:magfields}}
\end{figure}

Let us estimate how big the new quantum contribution, i.e. the last term in Eq.~(\ref{surface_new}), is. This term does not explicitly depend 
%\footnote{We do not discuss any origin of the field in Eq. (\ref{model}). We rely here on issues of Ref.\cite{MSS} where the dynamo for quadrupole poloidal field plus toroidal field non-vanishing at equator is preferred namely for fast rotating stars (like NS in our case!). This is contrary to the dynamo for the dipole poloidal field in the slow rotating Sun and toroidal field vanishing at the solar equator.}
on the NS rotation. We can take the axially symmetric magnetic field consisting of the toroidal and the quadrupole poloidal fields,
\begin{equation}\label{model}
  {\bf B}(r,\theta)= B_{p}(r)[\cos 2\theta {\bf e}_r + \sin 2\theta {\bf e}_{\theta}] +
  B_{\varphi}(r)\cos \theta {\bf e}_{\varphi},
\end{equation}
proposed by~\citet{MSS}. Here the angle $\theta$ is measured from the NS equator, that corresponds to $\theta=0$. Notice that the magnetic fields components are nonzero at the NS equator. The structure of this magnetic field is schematically shown in Fig.~\ref{fig:magfields}.

The factors $({\bf B}\cdot{\bf n})=B_r=B_p\cos 2\theta$ at the spherical surface where ${\bf n}={\bf e}_r$, and $B(R,\theta)=\sqrt{B_p^2 + B_{\varphi}^2\cos^2\theta}$ should be used
in the last term in Eq.~(\ref{surface_new}). Then, one obtains such a surface term in the form,
\begin{align}\label{quantum}
  \left(
    \frac{{\rm d}H}{{\rm d}t}
  \right)_\mathrm{quant} = &
  \frac{1}{\mu_e(R)}
  \oint_S B(R,\theta)({\bf B}\cdot {\bf n})\mathrm{d}^2S
  \nonumber
  \\
  & =-
  \left(
    \frac{2\pi R^2B_p}{\mu_e}
  \right)
  \int_{-1}^1(2x^2 - 1)\sqrt{B_p^2 + B_{\varphi}^2x^2}\mathrm{d}x
  \nonumber
  \\
  & = 
  \left(
    \frac{2\pi R^2B_pB_{\varphi}}{\mu_e}
  \right)
  \times A
  \sim
  \frac{B^2R^2}{\mu_e},
\end{align}
where $\mathrm{d}^2S=2\pi R^2\sin \theta \mathrm{d}\theta$. The factor $A$,
\begin{align}\label{factor}
  A(B_p/B_t) = & -\frac{1}{2}\left(\frac{B_p}{B_{\varphi}}\right)^2\sqrt{1 + \left(\frac{B_p}{B_{\varphi}}\right)^2} +\left(\frac{B_p}{B_{\varphi}}\right)^2\left[1 + \frac{1}{2}\left(\frac{B_p}{B_{\varphi}}\right)^2\right]
  \nonumber
  \\
  & \times
  \ln \left(\frac{B_{\varphi}}{B_p} + \frac{B_{\varphi}}{B_p}\sqrt{1 + \left(\frac{B_p}{B_{\varphi}}\right)^2}\right),
\end{align}
depends on the ratio $B_p/B_t$. For example, for $B_p\sim B_{\varphi}$ one gets $A\approx -0.05$. Thus, last estimate in Eq. (\ref{quantum}) is valid, $| ( \mathrm{d}H/ \mathrm{d}t)_\mathrm{quant}| \sim B^2R^2/\mu_e$.

The factor $A$ can be rewritten in the case, when $B_{\varphi}\gg B_{p}$,
\begin{equation}\label{factor2}
A=A(B_p/B_t)=\left(\frac{B_p}{B_{\varphi}}\right)^2\left[\ln \left(\frac{2B_{\varphi}}{B_p}\right) - \frac{1}{2}\right].
\end{equation}
In the opposite case, when $B_p\gg B_{\varphi}$, one gets
\begin{equation}\label{factor3}
\left(\frac{{\rm d}H}{{\rm d}t}\right)_\mathrm{quant}= \frac{4\pi R^2B_p^2}{3\mu_e},
\end{equation}
which can be obtained from Eq.~(\ref{quantum}).

We should compare the result in Eq.~(\ref{quantum}) with the classical surface term,
\begin{align}\label{classic}
  \left(
    \frac{{\rm d}H}{{\rm d}t}
  \right)_\mathrm{class}=&-\oint_S[({\bf B}\cdot{\bf A})({\bf v}\cdot{\bf n}) - ({\bf v}\cdot{\bf A})({\bf B}\cdot{\bf n})]\mathrm{d}^2S 
\nonumber\\
&= \oint (vA_{\varphi})({\bf B}\cdot{\bf n})\mathrm{d}^2S \sim  - B^2_pR^3\langle v\rangle,  
\end{align}
where we take the azimuthal rotation velocity ${\bf v}=v{\bf e}_{\varphi}$, i.e. the first term in Eq.~(\ref{classic}) is negligible, $({\bf v}\cdot{\bf n})=0$. The rotation velocity $v=\Omega R\cos \theta$ and $\langle v\rangle \approx \Omega R$ corresponds to the mean velocity after the integration over the polar angle in Eq.~(\ref{classic}). Notice that one can estimate the azimuthal potential $A_{\varphi}\sim - B_pR\sin 2\theta $
at the sphere $R< R_\mathrm{NS}$ for the poloidal component in Eq.~(\ref{model}) $B_{\theta}= - r^{-1}\partial_r(rA_{\varphi})=B_p\sin 2\theta$.

If we compare the classical and the quantum surface terms in the total sum,
\begin{equation}
  \left(
    \frac{{\rm d}H}{{\rm d}t}
  \right)=    \left(
    \frac{{\rm d}H}{{\rm d}t}
  \right)_\mathrm{class} +
  \left(
    \frac{{\rm d}H}{{\rm d}t}
  \right)_\mathrm{quant},
\end{equation}
we find that the quantum term is dominant only in the corotational reference frame, $\langle v\rangle - \Omega R= \delta v\ll 1$, namely for $R< \mu_e^{-1}/\delta v$. Using $R=10^5\,\text{cm}$ and $\mu_e^{-1}=2\times 10^{-13}\,\text{cm}$ for $\mu_e=100\,{\rm MeV}$, one obtains $\delta v < 10^{-18}$. This result correspond to the rigid rotation, $\partial_{\theta}\Omega=\partial_r\Omega=0$, i.e. $\Omega=\text{const}$. In this situation, $\langle v\rangle - \Omega R\to 0$. \citet{YakPet04} found that the neutron component in NS, which can be superfluid, has some deviations from the rigid rotation, $\delta v_n\neq 0$, contrary to the proton one, $\delta v_p=0$. Thus, our assumption on the absence of any differential rotation becomes invalid in such NSs. 

The result in Eq.~(\ref{quantum}) can be interpreted as an intertwining of the two thin tubes with magnetic fields having small base areas $S_p=\pi R_p^2$ and $S_t=\pi R_t^2$. Such tubes are situated at the sphere with radius $R\gg R_{p,t}$~\citep{Pri16},
\begin{equation}\label{quantum2}
  \left(
    \frac{{\rm d}H}{{\rm d}t}
  \right)_\mathrm{quant} =\dot{\theta}_\mathrm{pt}F_pF_t,
\end{equation}
where $F_t=B_tS_t$ and $F_p=B_pS_p$ are the fluxes of the magnetic field which tear off the two different toroids resulting from the toroidal and quadrupole poloidal components in Eq.~(\ref{model}). They penetrate the spherical surface $R<R_\mathrm{NS}$ floating up. The parameter
\begin{equation}\label{velocity}
\dot{\theta}_\mathrm{pt}=\frac{A}{\pi^2}\times \left(\frac{R}{R_p}\right)^2\left(\frac{1}{R_t^2\mu_e}\right)
\end{equation}
provides the angular velocity of the twisting of the magnetic loop bases one around other and causing the interlacing of the flux tubes. For instance, one can evaluate the parameter in Eq.~\eqref{quantum2} as $\dot{\theta}_\mathrm{pt}\sim 10^7 \times (6A/\pi^2)\,\text{s}^{-1}$ for $R_p=R_t= 1\,\text{cm}$, $R=10^5\,\text{cm}$. Thus, within the time $t\sim 3\times 10^{-6}\,\text{s}\gg 10^{-12}\,\text{s}$,\footnote{We substitute for estimates $A=-0.05$ for $B_p=B_t$ in Eq. (\ref{factor}) when already $n_\mathrm{R}=n_\mathrm{L}$. It means that the CME is irrelevant in NS, $\mu_5=0$.} the magnetic helicity evolution in Eq.~(\ref{velocity}) leads to a flux tangling with the linkage (topology) number $L_\mathrm{pt}=\int \dot{\theta}_\mathrm{pt}dt=L_{12}\sim 1$ in the famous Gauss formula $H= 2L_{12}F_1F_2$ where $L_{12}=L_\mathrm{pt}= 1$ is conserved afterwards~\citep{Pri16}.

\subsection{Magnetic helicity flow through the equator of a rotating NS\label{sec:APPL}}

Now we turn to the problem of the magnetic helicity evolution in a rotating NS accounting for the electroweak interaction between fermions. For this purpose, we use Eq.~(\ref{eq:n5gen}). We mentioned in Sec.~\ref{sec:ELCURR} that the spin-flip rate is huge in NS. Hence, $\dot{n}_5\to0$ in Eq.~\eqref{eq:n5gen}. Using Eq.~\eqref{eq:n5gen}, we get that the chiral imbalance reaches the saturated value, 
\begin{equation}\label{eq:n5sat}
  n_5 \to n_5^{(\text{sat})}=\frac{e^{2}}{2\pi^{2}V\Gamma}\int
  \left[
    \left(
      \mathbf{E}\cdot\mathbf{B}
    \right)-
    \frac{1}{e}
    \left(
      \mathbf{E}\cdot\nabla\times\mathbf{V}_{5}
    \right)
  \right]
  \mathrm{d}^{3}x,
\end{equation}
We shall demonstrate below that we can neglect $n_5^{(\text{sat})}$ in an old star. Thus, the saturated densities of left and right fermions are equal.

If both $n_{5}$ and $\dot{n}_{5}$ vanish in Eq.~\eqref{eq:n5gen}, then, taking into account the fact that
\begin{equation}
  \frac{\mathrm{d}H}{\mathrm{d}t}=-2\int
  \left(
    \mathbf{E}\cdot\mathbf{B}
  \right)
  \mathrm{d}^{3}x,
\end{equation}
where
\begin{equation}\label{eq:maghel}
  H=\int
  \left(
    \mathbf{A}\cdot\mathbf{B}
  \right)
  \mathrm{d}^{3}x,
\end{equation}
is the magnetic helicity, we obtain the contribution of the electroweak interaction to the evolution of the magnetic helicity in the form,
\begin{equation}\label{eq:helevol}
  \left(
    \frac{\mathrm{d}H}{\mathrm{d}t}
  \right)_\mathrm{EW} =
  -2\frac{V_{5}}{e}\int
  \left(
    \mathbf{E}\cdot\bm{\omega}
  \right)
  \mathrm{d}^{3}x=-2\frac{V_{5}}{e\sigma}\int
  \left(
    \nabla\times\mathbf{B}\cdot\bm{\omega}
  \right)
  \mathrm{d}^{3}x.
\end{equation}
Here we account for the Maxwell equations and the fact that
$\mathbf{E}=\mathbf{J}/\sigma=(\nabla\times\mathbf{B})/\sigma$, where
$\mathbf{J}$ is the ohmic electric current and $\sigma$ is the conductivity. Note that the analogous consideration was made in Sec.~\ref{sec:HELEVOL}.

Equation~\eqref{eq:helevol} should be included to the contribution of the classical MHD for the magnetic helicity behavior (see Eq.~\eqref{eq:helevolclass}),
\begin{equation}\label{eq:helevolclass}
  \left(
    \frac{\mathrm{d}H}{\mathrm{d}t}
  \right)_\mathrm{class} =
  -\frac{2}{\sigma}\int
  \left(
    \mathbf{J}\cdot\mathbf{B}
  \right)
  \mathrm{d}^{3}x+
  2\oint
  \left[
    ( \mathbf{B}\cdot\mathbf{A} ) ( \mathbf{v}\cdot\mathbf{n} ) -
    ( \mathbf{v}\cdot\mathbf{A} ) ( \mathbf{B}\cdot\mathbf{n} )
  \right]
  \mathrm{d}^{2}S,
\end{equation}
where $\mathbf{n}$ is the external normal to the surface of a star. The total rate of the helicity variation is $\dot{H} = (\dot{H})_\mathrm{class} + (\dot{H})_\mathrm{EW}$, where the electroweak and the classical MHD contributions are present in Eqs.~\eqref{eq:helevol} and~\eqref{eq:helevolclass}. We omit the subscript EW since we are interested mainly in the $(\dot{H})_\mathrm{EW}$ term here.

Equation~(\ref{eq:helevol}) represents the new mechanism for the
helicity evolution due to the electroweak interaction of chiral
electrons with an inhomogeneous background matter. The inhomogeneity of
the interaction of chiral fermions is present in the dependence of $\mathbf{V}_{\mathrm{L,R}}=V_{\mathrm{L,R}}^{0}\mathbf{v}$
on spatial coordinates. It can be implemented if we consider a rotating matter
with the velocity $\mathbf{v}(\mathbf{x})=\bm{\omega}\times\mathbf{x}$.
The anomalous contribution to the helicity change in Eq.~(\ref{eq:helevol})
takes place if $(\nabla\times\mathbf{B})\parallel\bm{\omega}$. It can be the case in a rotating NS having a toroidal component
of the magnetic field.
%It should be noted that \citet{DvoSem18b} studied the influence of quantum
%effects on the magnetic helicity evolution, analogous to that in Eq.~(\ref{eq:helevol}).

\begin{figure}
  \centering
  \includegraphics[scale=0.65]{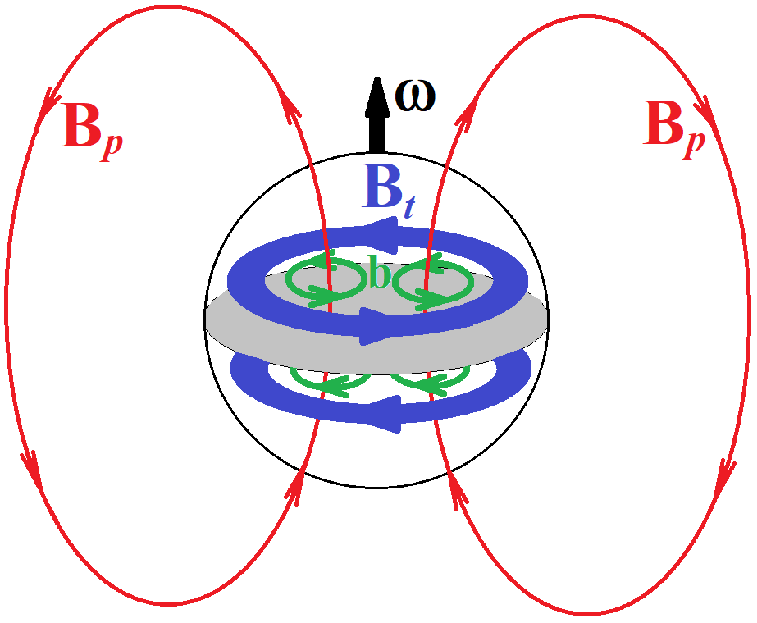}
  \caption{The schematic structure of the considered stellar magnetic field.
  The toroidal field $\mathbf{B}_{t}$
  is placed within two blue tori with opposite directions above
  and below the equator of a star, which is filled by the grey color. 
  The poloidal field $\mathbf{B}_{p}$, which can nonzero outside
  the star, is depicted by red lines. The fluctuations
  of the magnetic field are shown with small green rings. They can exist in both
  hemispheres of a star. The direction of the magnetic
  field $\mathbf{b}$ in such rings coincides with that for the toroidal component.
  The figure is taken from the work by~\citet{Dvo20}.\label{fig:starB}}
\end{figure}

We suppose that a compact star possesses the dipole configuration of the
magnetic field: the two tori
of the toriodal magnetic field $\mathbf{B}_{t}$ and the poloidal field component $\mathbf{B}_{p}$. The directions of the field in the tori are different in the 
opposite hemispheres of a star. It is schematically depicted in Fig.~\ref{fig:starB}.
\citet{BraNor06} showed that a stellar magnetic field, which has only either a toroidal or a poloidal 
component, is unstable; cf. Fig.~\ref{fig:magfields}.

Let us assume that the toroidal field in one of the hemispheres
of the star produces magnetic field rings or vortices, which
disattach from the corresponding torus. \citet{And07} found that certain turbulence processes,
which occur inside some compact rotating stars, can result in the formation of such rings. These
vortices are depicted in Fig.~\ref{fig:starB} with green color. We notice that the direction of the magnetic field in a ring coincides
with that in a torus, in which the toroidal component is concentrated.
The electroweak mechanism, given in Eq.~\eqref{eq:helevol},
acts on these vortices.

The described process results in the modification
of the magnetic field of a star. Namely, the magnetic helicity starts to flow through the stellar equator. Indeed, there are different signs
of $\dot{H}$ in opposite hemispheres since $\mathbf{B}_{t}$, as well as rings,
has opposite directions in the southern and northern hemispheres.
Thus, the total helicity is conserved, $\dot{H}_{\mathrm{tot}}=\dot{H}_{\mathrm{S}}+\dot{H}_{\mathrm{N}}=0$.
Here $H_{\mathrm{S,N}}$ are the helicities of the southern and northern
hemispheres.

We take that the mean strength of the magnetic field in a ring is $\mathbf{b}$
and the mean width of a flux tube of a vortex is $r$. Assuming that all the magnetic field in the northern
hemisphere modifies into rings and taking into account the conservation of the magnetic
flux, we obtain that $br^{2}N=B_{t}R_{t}^{2}$, where
$N$ is the total number of rings and $R_{t}$ is the torus radius. Thus, we can evaluate $(\nabla\times\mathbf{B})$ in Eq.~(\ref{eq:helevol})
as
\begin{equation}
  (\nabla\times\mathbf{B})\sim N\frac{b}{r}\sim\frac{B_{t}R_{t}^{2}}{r^{3}}.
\end{equation}
Let us consider, for instance, the northern hemisphere. Using Eq.~(\ref{eq:helevol}),
we have the rate of the helicity change in the form, 
\begin{equation}
  \dot{H}_{\mathrm{N}} \sim
  -\frac{4}{3}\pi B_{t}R_{t}^{2}\omega\frac{V_{5}}{e\sigma}\frac{R^{3}}{r^{3}},
\end{equation}
where $R\sim10\,\text{km}$ is the radius of NS. The typical time of
the helicity change inside of the hemisphere is
\begin{equation}\label{eq:relaxtime}
  \tau\sim\frac{H_{\mathrm{N}}}{|\dot{H}_{\mathrm{N}}|}=
  \frac{3}{4\pi}\frac{e B_{p}\sigma r^{3}}{\omega V_{5}R}.
\end{equation}
\citet{Ber99} proposed the  alternative definition of the helicity compared to that in Eq.~(\ref{eq:maghel}): $H=2L\Phi_{1}\Phi_{2}$, where
$\Phi_{1,2}$ are the magnetic fluxes linked and $L=0,\pm1,\dotsc$
is the linkage number (see also Sec.~\ref{sec:HELEVOL}). Such a definition is used in Eq.~\eqref{eq:relaxtime}. We take
that $H_{\mathrm{N}}\sim B_{t}R_{t}^{2}B_{p}R^{2}$ is the total helicity
in the northern hemisphere in Eq.~(\ref{eq:relaxtime}).

We apply the mechanism for the helicity change, proposed above, inside NS. \citet{DvoSem15} obtained that $V_{5}=G_{\mathrm{F}}n_{n}/2\sqrt{2}=6\,\text{eV}$ in this case. Here $n_{n}=1.8\times10^{38}\,\text{cm}^{-3}$ is the density of neutrons
in NS. The possibility to use the chiral phenomena for relativistic electrons in dense
matter of NS is justified by the fact that the Fermi momentum of such particles $p_{\mathrm{F}e}=(3\pi^{2}n_{e})^{1/3}\sim10^{2}\,\text{MeV}$ is much greater than their masses. Here we take that the density of electrons $n_{e}$ is about 5\% of $n_{n}$
inside NS.

We suppose that $B_{p}=10^{12}\,\text{G}$ is the typical magnetic field of a pulsar, $\omega=10^{3}\,\text{s}^{-1}$ is the angular velocity
of a millisecond pulsar. \citet{SchSht18} found that the electric
conductivity of the NS matter having the temperature $T\sim10^{7}\,\text{K}$ is $\sigma=10^{6}\,\text{GeV}$.
\citet{GusDom16} predicted the existence of the magnetic flux tubes with $r\sim6\times10^{-5}\,\text{cm}$ in background matter of certain NSs.
We obtain that $\tau=1.3\times10^{11}\,\text{s}=4\times10^{3}\,\text{yr}$ if we use the above values in Eq.~(\ref{eq:relaxtime}).

We describe the new possibility for the magnetic helicity to flow through the equator of NS
driven by the electroweak interaction of chiral fermions with the
rotating matter. \citet{Sor13} found that the helicity flux through the equator
is related to the periodic
variation of the stellar magnetic field. For instance, it can be the
known solar cycle with the period of $22\,\text{yr}$. Therefore we
can conclude that the obtained $\tau\sim10^{3}\,\text{yr}$ is related to
a periodic variation of the magnetic field in pulsars. We notice
that \citet{Con07} observed analogous cycles in certain NSs with $10^2\,\text{yr}<\tau<10^{4}\,\text{yr}$ by studying the pulsars spin-down. The mechanism predicted in the present work is a possible explanation of the observations of~\citet{Con07}.

Now, we show that we can omit $n_5^{(\text{sat})}$ in Eq.~\eqref{eq:n5sat}. It was mentioned above that the total magnetic helicity of NS is conserved. The second term in Eq.~\eqref{eq:n5sat}, $\sim \left( \mathbf{E}\cdot\nabla\times\mathbf{V}_{5}   \right)$, is important while turbulent vortices of the magnetic field are produced. Thus we can compute the contribution to $n_5^{(\text{sat})}$ at the beginning of the process of the helicity change, that is from the first term in the integrand, $\sim \left( \mathbf{E}\cdot\mathbf{B} \right)$.
Accounting for that $\mathbf{E}=(\nabla\times\mathbf{B})/\sigma$, we get that $n_5^{(\text{sat})}\sim e^{2}B^2/2\pi^{2}R\sigma\Gamma$.
If one takes the typical magnetic field of a pulsar $B = 10^{12}\,\text{G}$, the radius of NS $R=10\,\text{km}$, the rate of the spin-flip $\Gamma = 10^{11}\,\text{s}^{-1}$, and the electric conductivity $\sigma=10^{6}\,\text{GeV}$, we obtain that $n_5^{(\text{sat})}\sim 10^{12}\,\text{cm}^{-3}$.

We should compare the value of $n_5^{(\text{sat})}$ estimated with the initial chiral imbalance $n_5(0)$ which is produced at the formation of NS. \citet{DvoSem15} proposed that a seed chiral imbalance can be created in the parity violating direct Urca processes leading to the washing out of left electrons from NS. The energy scale of Urca processes is $\mu_5(0) = m_n - m_p \sim 1\,\text{MeV}$, where $\mu_5(0) = (\mu_\mathrm{R} - \mu_\mathrm{L})/2 = \pi^2 n_5(0)/p_{\mathrm{F}e}^2$ and $m_{p,n}$ are the masses of a proton and a neutron.
\citet{DvoSem15,Dvo16a,DvoSemSok20} used this value of $\mu_5(0)$ to study the problem of magnetars with help of chiral phenomena. Supposing that $p_{\mathrm{F}e}\sim10^{2}\,\text{MeV}$, one obtains that $n_5(0) \sim 10^{33}\,\text{cm}^{-3}$. Therefore, one can see that $n_5^{(\text{sat})} \ll n_5(0)$. Hence, we are able to neglect both $n_5$ and $\dot{n}_5$ in Eq.~\eqref{eq:n5gen} in matter of an old NS.

Nevertheless one can have a backreaction from the magnetic field to the chiral imbalance. \citet{DvoSemSok20} studied this process recently and  found that there is a spike in the dependence of $n_5$ on time associated with the change of the polarity of a young NS by considering the full set of the AMHD equations. Concerning the problem discussed in our work, $n_5$ can grow sharply within the time $\tau$ evaluated in Eq.~\eqref{eq:relaxtime}. However, one needs a separate study to clarify this issue.

\section{Conclusion\label{sec:CONCL}}

In the present work, we have studied two different possibilities for the evolution of the magnetic helicity in a compact star. First, we have derived the new Eq.~(\ref{new_law}) accounting for  the new quantum surface term in Eq.~(\ref{quantum_correction}) with the spin effects in plasma by averaging the Adler anomaly in Eq.~(\ref{pseudovector}). This new term is important for the magnetic helicity evolution at the spherical surface  around a finite volume of a dense NS.

The  evolution of the magnetic helicity could potentially cause the the magnetic field lines reconnection at the surface and result in flares from outer boundary of a star taking place in the stellar magnetosphere. We did not solve such a problem in the present work. We are trying just to find how strong the new magnetization effect should be deeply within NS core where our approximations are still valid. For this purpose we had to consider small base areas $L^2=(R\Delta \theta)^2\ll R^2$ at the surface with the radius $R< R_\mathrm{NS}$ for the corresponding thin magnetic tubes intersecting such a surface for both  poloidal and toroidal components in Eq. (\ref{model}). It happens since both the magnetic helicity diffusion $-2\int_V \mathrm{d}^3x ({\bf E}\cdot{\bf B})$ and the new quantum term in Eq.~(\ref{dissipation}) are important only at small scales, $L\ll R$, when the evolution time is greater than a long diffusion time for the high conductivity of matter in NS, $t > \tau_\mathrm{D}=L^2\sigma_\mathrm{cond}$. Notice that the contribution of the new term is not compared with the diffusion losses since we consider above an ideal plasma in the limit $\sigma_\mathrm{cond}\to \infty$.

We find that the inequalities $m_e^2\ll eB\ll p_{\mathrm{F}_e}^2$ are fulfilled in magnetars with strong magnetic fields $B\sim 10^{15}\,{\rm G}$.
Moreover, one has that $p_{\mathrm{F}_e}=100\,{\rm MeV}$ within a core of these NSs with ultrarelativistic electrons. Thus, the CME contribution occurs small due to a small population of electrons at the main Landau level $n=0$; cf. Eq.~(\ref{real}). Nevertheless, namely these electrons provide the magnetic helicity diffusion through the new quantum contribution in the evolution Eq.~(\ref{standard3}). 

Note that the WKB approximation, with $n\gg 1$, in Eq.~(\ref{real}), when paramagnetic (spin) contribution is a small correction in the Landau spectrum Eq.~(\ref{levels}), simplifies the derivation of the pseudoscalar term $2\mathrm{i}m_e\langle \bar{\psi}\gamma_5\psi\rangle=\mu_\mathrm{B}^{-1}(\nabla\cdot \bm{\mathcal{S}}_5)$ made by~\citet{DvoSem18b}. 

The WKB approximation is invalid in the NS crust, where degenerate electrons become nonrelativistic, $eB\gg m_e^2\gg p_{\mathrm{F}_e}^2$. It means that they populate only the main Landau level $n=0$. In this situation, one can calculate the pseudovector $\bm{\mathcal{S}}_5$ in such a medium. It is expected to be comparable with the standard magnetization in Eq.~(\ref{NRplasma}). This case should be of special interest since it corresponds to the outer NS surface where the magnetic helicity evolution can lead to gamma or X-ray bursts happening in magnetars.

The second task, considered in the present work, is related to the chiral phenomena in spatially inhomogeneous
matter accounting for the electroweak interaction between chiral electrons and
background fermions. In Sec.~\ref{sec:EVOL}, we have derived
the wave equation for chiral electrons electroweakly interacting a background matter having a nonuniform density and an
arbitrary velocity. We have analyzed this equation basing on the concept of the Berry phase. In Sec.~\ref{sec:EVOL}, we have derived the kinetic equations for left and right particles, as well as the effective
actions which generalize the results of~\citet{DvoSem17,DvoSem18a}.

Then, in Sec.~\ref{sec:ELCURR}, we have derived the corrections to the Adler-Bell-Jackiw anomaly
and to the anomalous current from the
electroweak interaction of chiral electrons with inhomogeneous matter.
In this situation, we have considered a particular example of the rotating matter.
The obtained contribution to the anomalous current is consistent with the
result of~\citet{Dvo15}.

We have applied our results
for the description of the evolution of magnetic fields in dense matter
of a compact star in Sec.~\ref{sec:APPL}. We have obtained the contribution to the magnetic helicity evolution
of a rotating NS due to the electroweak
interaction of chiral fermions with background neutrons by supposing that the chiral imbalance vanishes
in collisions between particles. This effect
generates the magnetic helicity flux through the equator of a compact star.
We have assumed that magnetic vortices in the form of
rings are formed because of the magnetic turbulence in NS. Then, we have computed the
typical time of the change of the magnetic helicity in one of the hemispheres
of a compact star. We have demonstrated that the total magnetic helicity of NS is constant.

\citet{Sor13} showed that the flux of the magnetic helicity through the stellar equator is related to the periodic cycle of the magnetic activity of a star analogous to the known solar cycle. The typical time, obtained
in Sec.~\ref{sec:APPL}, is compared to the period of the cyclic
electromagnetic activity of certain pulsars observed by~\citet{Con07}.
Therefore, the mechanism developed in this work is a feasible explanation
of the results of~\citet{Con07}.

\end{document}